\documentclass[12pt]{iopart}

\usepackage{graphicx,setspace}
\usepackage{dsfont}
\usepackage{amssymb}
\usepackage{mathrsfs}
\usepackage{psfrag}
\usepackage{cases}

\begin{document}
\title{Quantum dynamics of the damped harmonic oscillator}
\author{T G Philbin}
\address{School of Physics and Astronomy, University of St Andrews,
North Haugh, St Andrews, Fife KY16 9SS,
Scotland, UK.}
\ead{tgp3@st-andrews.ac.uk}

\begin{abstract}
The quantum theory of the damped harmonic oscillator has been a subject of continual investigation since the 1930s. The obstacle to quantization created by the dissipation of energy is usually dealt with by including a discrete set of additional harmonic oscillators as a reservoir. But a discrete reservoir cannot directly yield dynamics such as Ohmic damping (proportional to velocity) of the oscillator of interest. By using a continuum of oscillators as a reservoir, we canonically quantize the harmonic oscillator with Ohmic damping and also with general damping behaviour. The dynamics of a damped oscillator is determined by an arbitrary effective susceptibility that obeys Kramers-Kronig relations. This approach offers an alternative description of nano-mechanical oscillators and opto-mechanical systems.
\end{abstract}


\section{Introduction}
Few classical dynamical systems are as simple or important as the one-dimensional damped harmonic oscillator:
\begin{equation} \label{qdampf}
\ddot{q}+\gamma \dot{q}+\omega_0^2 q=f(t).
\end{equation}
But the simplicity of this dynamical system does not survive the transition to quantum mechanics. The presence of dissipation in (\ref{qdampf}) (or amplification with $t\rightarrow -t$) leads to severe difficulties with its quantization, a problem that has attracted repeated investigation from the 1930s to the present day (some historical information is given in \cite{dek81,um02,wei08}). A brief account of the different approaches to quantizing (\ref{qdampf}) will explain the rather straightforward, but crucial, respect in which the starting point of this paper differs from previous work.

The central obstacle in quantizing the dynamics (\ref{qdampf}) is that a Hamiltonian is required that will generate the quantum time evolution (a Lagrangian will suffice for the path-integral approach). A Lagrangian and Hamiltonian exist that give the equation of motion (\ref{qdampf}), but they are time dependent because of the dissipation (or amplification) and this leads to  difficulties in implementing the canonical commutation relation~\cite{dek81,um02}. If the dissipated energy is included through extra dynamical degrees of freedom, then the problem of a time-dependent Hamiltonian can be avoided. The simplest approach is to write a two-body Hamiltonian describing one damped and one amplified oscillator, with conserved total energy~\cite{bat31}. But the canonical variables of this system~\cite{bat31} are not the positions and momenta of the two oscillators, so again there is no straightforward way of imposing canonical commutation relations on each oscillator~\cite{dek81,um02,bla02,lat05,bal11,maj12}. In recent times the most popular approach is to treat a damped harmonic oscillator as a free oscillator coupled to a reservoir of oscillators of different frequencies. The apparently universal practice  for investigations of the damped harmonic oscillator has been to use a discrete set of oscillators for the reservoir.\footnote{There are of course countless other contexts in which use is made of a discrete reservoir.} The resulting form of the Hamiltonian is attributed to Magalinskii~\cite{mag59}, and it is also the most popular starting point for attempts to describe quantum Brownian motion (with a free particle coupled to the reservoir)~\cite{wei08,fey63,cal83,smi90,han05,han08,ing09a,ing09b,dat10,ing12}. This approach does not give the dynamics (\ref{qdampf}) for the oscillator of interest, with a damping proportional to velocity; instead, the oscillator equation of motion can be written with a term $\int_{t0}^tds\,g(t-s)\dot{q}(s)$, where $g(t)$ is an integral kernel dependent on the coupling to the reservoir~\cite{wei08}. In order to describe damping proportional to velocity, often called Ohmic damping in view of the electrical application of (\ref{qdampf}), a limiting procedure $g(t)\rightarrow 2\gamma\delta(t)$ must be employed at some point~\cite{wei08}. The limit that produces Ohmic damping (or amplification) involves arbitrarily decreasing the frequency spacing between oscillators in the reservoir while maintaining reservoir oscillators of arbitrarily high frequencies; in other words, the reservoir is effectively regarded as a continuum, but only after the dynamical equations have been solved under the assumption that the reservoir is a discrete set. The impossibility of achieving Ohmic damping with a discrete reservoir and the emergence of Ohmic damping through a delicate continuum limit, after the dynamics has been solved, is discussed in great detail by Tatarskii~\cite{tat87}. In addition to the approaches just described, which seek to employ the standard quantization rules, there are phenomenological approaches to the damped harmonic oscillator where no rigorous quantization is attempted (reference~\cite{gra84} is just one example of such approaches).

The starting point of the results presented here is a harmonic oscillator coupled to a reservoir, where the latter is a continuum of oscillators of all positive frequencies. Use of a continuum reservoir from the outset gives a much richer dynamical system compared to the use of a discrete reservoir. As the system we will be analyzing has an uncountable number of degrees of freedom, it acquires many of the properties of a field theory, and is qualitatively different from a countable set of coupled oscillators, even if the latter set is infinite. We will describe the general method for solving the dynamical system with a continuum reservoir and solve exactly the case of Ohmic damping and amplification, which will emerge from a particular choice of coupling to the reservoir. The quantization for general damping and Ohmic damping is treated in detail, including the diagonalization of the Hamiltonian and the case of thermal equilibrium.

The continuum reservoir appears to originate with Huttner and Barnett~\cite{hut92}, a paper that is very well known and yet whose technical innovation and importance have been under-appreciated. An example of what can be achieved with a continuum reservoir is the canonical quantization of the macroscopic Maxwell equations for arbitrary media obeying the Kramers-Kronig relations~\cite{phi10,phi11}, including bi-anisotropic and moving media~\cite{hor11,hor12}. In the present paper the continuum reservoir will again show its power by allowing an exact treatment of the dynamics (\ref{qdampf}) and its canonical quantization. More importantly, the continuum reservoir will naturally lead to the characterization of a quantum damped harmonic oscillator by an arbitrary effective susceptibility obeying Kramers-Kronig relations. Because of this last property, the quantum damped harmonic oscillator will be found to have much in common with the quantum theory of light in macroscopic media. This offers a alternative framework for describing macroscopic quantum oscillators, which are now a subject of some remarkable experiments~\cite{oco10,teu11,cha11,asp10,poo12}. Rather than attempting to capture the immensely complicated microscopic physics, the results in this paper suggest that macroscopic quantum oscillators may be describable by effective susceptibilities that are to be experimentally measured, just as the electromagnetic properties of macroscopic media are captured by measured permittivities and permeabilities.

The price to be paid for employing the continuum reservoir is the extra mathematical complexity compared to the discrete case. Many aspects of this mathematical apparatus, which provides an important and unusual addition to standard quantum field theory, have still not been fully explored. As well as treating the specific problem of the damped harmonic oscillator, the results presented here give further insight into the remarkably rich classical and quantum physics of a continuum reservoir.

Section~\ref{sec:L} gives the Lagrangian and equations of motion of an oscillator coupled to a continuum reservoir. In sections~\ref{sec:solnI} and~\ref{sec:solnII} the dynamics for a particular coupling that gives damping proportional to velocity is solved in detail. The system is quantized in section~\ref{sec:quantization} and the diagonalization of the Hamiltonian is addressed. In sections~\ref{sec:coh} and~\ref{sec:thermal} coherent-state solutions and thermal equilibrium are treated. Possible applications and extensions of the results are discussed in section~\ref{sec:concl}.

\section{Lagrangian and dynamical equations}  \label{sec:L}
We consider the dynamical system with Lagrangian 
\begin{equation}  \label{L}
L=\frac{1}{2}\dot{q}^2-\frac{1}{2}\omega_0^2 q^2+\frac{1}{2}\int_0^\infty\rmd\omega\left(\dot{X}_\omega^2-\omega^2X_\omega^2\right)+\int_0^\infty\rmd\omega\,\alpha(\omega)qX_\omega.
\end{equation}
This describes a harmonic oscillator with displacement $q$ and frequency $\omega_0$ and a reservoir of oscillators with displacements $X_\omega$ and frequencies $\omega\in[0,\infty)$, with the $q$-oscillator linearly coupled to the reservoir by an arbitrary coupling function $\alpha(\omega)$. We couple only displacements in (\ref{L}), not velocities, and the reservoir oscillators are not directly coupled to each other. The variables $q(t)$ and $X_\omega(t)$ are functions only of time, so the entire system may be viewed as located at one point in space. The dynamical variables in (\ref{L}) are thus not \emph{fields} in the conventional sense, but the dependence of  $X_\omega(t)$ on the continuous quantity $\omega$ will give this system many of the properties of a field theory, in sharp contrast to the case of a discrete reservoir. The total energy of the system described by (\ref{L}) is
\begin{equation}   \label{E}
E=\frac{1}{2}\dot{q}^2+\frac{1}{2}\omega_0^2 q^2+\frac{1}{2}\int_0^\infty\rmd\omega\left(\dot{X}_\omega^2+\omega^2X_\omega^2\right)-\int_0^\infty\rmd\omega\,\alpha(\omega)qX_\omega.
\end{equation}

The system (\ref{L}) is close to being a drastic simplification of the Huttner-Barnett model~\cite{hut92} of a dielectric coupled to the electromagnetic field: if the electromagnetic field is removed along with the spatial dependence of the medium and the reservoir, then the Huttner-Barnett model almost reduces to the system (\ref{L}), the difference being that the $q$-oscillator would be coupled to $\dot{X}_\omega$ rather than  $X_\omega$. The relationship of our system to the Huttner-Barnett model will be commented on at several points, as some of the technical achievements of Huttner-Barnett will be closely related to results here. The main differences from Huttner-Barnett, in addition to the simplifications just described, are that (i) we do not rely solely on a retarded or advanced solution of the reservoir dynamics, (ii) much of our time will be spent in obtaining the exact solution for a specific coupling function, whereas Huttner and Barnett consider only a general coupling function, and (iii) our coupling term is of a different form.

The Euler-Lagrange equations of (\ref{L}) are
\begin{eqnarray}
\ddot{q}+\omega_0^2 q-\int_0^\infty\rmd\omega\,\alpha(\omega)X_\omega=0,  \label{qeq} \\
\ddot{X}_\omega+\omega^2 X_\omega-\alpha(\omega)q=0. \label{Xeq}
\end{eqnarray}
The general solution of the reservoir equation (\ref{Xeq}) can be written
\begin{equation}   \label{Xgensol}
\fl
\eqalign{ X_\omega(t)=A_0(\omega)\cos\omega t+B_0(\omega)\sin\omega t+\frac{\alpha(\omega)}{\omega}\int_0^t\rmd t' \, q(t') \sin\left[\omega(t-t')\right]\,e^{-0^+|t-t'|},  \cr
\qquad\qquad\qquad A_0(\omega)=X_\omega(0), \qquad B_0(\omega)=\frac{1}{\omega}\dot{X}_\omega(0), }
\end{equation}
where $0^+$ is a positive infinitesimal quantity. By means of (\ref{Xgensol}) we can impose arbitrary displacements $X_\omega(0)$ and velocities $\dot{X}_\omega(0)$ on the reservoir at $t=0$ (these are not ``initial" conditions because the solution (\ref{Xgensol}) is valid for all $t$). The choice of $t=0$ is of course arbitrary, but no generality is lost by the form (\ref{Xgensol}) if we wish to impose conditions on the reservoir at some finite time (the imposition of conditions in the infinite past or future is dealt with later in this section). The presence of the exponential in (\ref{Xgensol}) is important for taking the Fourier transform of this general solution, and it can be understood as follows. Solutions of (\ref{Xeq}) can be constructed using a Green function $G(t)$ defined by
\begin{equation} \label{Geq}
\ddot{G}+\omega^2 G=\delta(t).
\end{equation}
For example, the retarded $G_r(t)$ and advanced $G_a(t)$ Green functions are
\begin{eqnarray}
G_r(t)=\frac{1}{\omega}\theta(t)\sin\omega t \,e^{-0^+t} ,  \label{Gr} \\
G_a(t)=-\frac{1}{\omega}\theta(-t)\sin\omega t \,e^{0^+t} , \label{Ga}
\end{eqnarray}
where $\theta(t)$ is the step function. The exponential factors are necessary in the Green functions (\ref{Gr}) and (\ref{Ga}) for their Fourier transforms to exist, and the infinitesimal number $0^+$ gives the familiar pole prescriptions in the frequency domain, with $G_r(\omega)$ analytic in the upper-half complex $\omega$-plane and $G_a(\omega)$ analytic in the lower-half plane. The general solution (\ref{Xgensol}) is constructed with the difference
\begin{equation}  \label{G-}
G_r(t)-G_a(t)=\frac{1}{\omega}\sin\omega t \,e^{-0^+|t|},
\end{equation}
which is a solution of the homogeneous version of (\ref{Geq}) (i.e.\ (\ref{Geq}) without the delta function). This gives the exponential factor in (\ref{Xgensol}) that is required to define the Fourier transform of $X_\omega(t)$.

To complete the solution for the dynamics we must substitute (\ref{Xgensol}) into (\ref{qeq}) and solve the resulting equation for $q(t)$:
\begin{eqnarray}
\fl
\ddot{q}+\omega_0^2 q-\int_0^\infty\rmd\omega\int_0^t\rmd t'\,q(t')\frac{\alpha^2(\omega)}{\omega} \sin\left[\omega(t-t')\right]\,e^{-0^+|t-t'|} \nonumber \\
-\int_0^\infty\rmd\omega\,\alpha(\omega)\left\{\frac{1}{2}\left[A_0(\omega)+\rmi B_0(\omega)\right]\exp(-\rmi\omega t)+\mathrm{c.c.}\right\}=0.  \label{qeqeff}
\end{eqnarray}
For most coupling functions $\alpha(\omega)$ this integro-differential equation is difficult to solve in the time domain. When written in the frequency domain however, (\ref{qeqeff}) becomes an integral equation that can be solved by a systematic procedure once the coupling function $\alpha(\omega)$ is chosen. Instead of directly Fourier transforming (\ref{qeqeff}), it is a little simpler to Fourier transform (\ref{Xgensol}) and substitute the result into the Fourier transform of (\ref{qeq}). Defining the relation
\begin{equation} \label{Four}
f(t)=\frac{1}{2\pi}\int_{-\infty}^{\infty}\rmd \omega\,f(\omega)\exp(-\rmi \omega t)
\end{equation}
between the time and frequency domains, we obtain from (\ref{Xgensol})
\begin{eqnarray}
\fl   X_\omega(\omega')=\pi\left[A_0(\omega)+\rmi B_0(\omega)\right]\delta(\omega-\omega')+\pi\left[A_0(\omega)-\rmi B_0(\omega)\right]\delta(\omega+\omega')
 \nonumber \\
+\mathrm{P}\frac{\alpha(\omega)q(\omega')}{\omega^2-{\omega'}^2}+ \frac{\alpha(\omega)}{2\omega}\left[\delta(\omega-\omega')-\delta(\omega+\omega')\right]\mathrm{P}\int_{-\infty}^{\infty}\rmd \xi\frac{q(\xi)}{\xi-\omega'},       \label{Xfreqgensol}
\end{eqnarray}
where the infinitesimal quantity $0^+$ in (\ref{Xgensol}) has given rise to principal-value terms denoted with a P (delta-function terms arising from $0^+$ are found to cancel). The Fourier transforms of the basic equations (\ref{qeq}) and (\ref{Xeq}) are
\begin{eqnarray} 
(-{\omega'}^2+\omega_0^2)q(\omega')-\int_{0}^{\infty}\rmd \omega\,\alpha(\omega)X_\omega(\omega')=0,   \label{qeqfreq}  \\
(-{\omega'}^2+\omega^2)X_\omega(\omega')-\alpha(\omega)q(\omega')=0.   \label{Xeqfreq}
\end{eqnarray}
One verifies that (\ref{Xfreqgensol}) solves (\ref{Xeqfreq}), and substitution of (\ref{Xfreqgensol}) in (\ref{qeqfreq}) yields an integral equation for $q(\omega')$:
\begin{eqnarray}
\fl
\left[{\omega'}^2-\omega_0^2+\mathrm{P}\int_{0}^{\infty}\rmd \omega\frac{\alpha^2(\omega)}{\omega^2-{\omega'}^2}\right]q(\omega')+\frac{\alpha^2(|\omega'|)}{2\omega'}\mathrm{P}\int_{-\infty}^{\infty}\rmd \xi\frac{q(\xi)}{\xi-\omega'}   \nonumber \\
=-\pi\alpha(|\omega'|)\left[A_0(|\omega'|)+\rmi \,\mathrm{sgn}(\omega')B_0(|\omega'|)\right]  \label{qinteq}
\end{eqnarray}
Equation (\ref{qinteq}) is consistent with the relation $q^*(\omega')=q(-\omega')$, which holds because $q(t)$ is real. Once $q(\omega')$ is found from (\ref{qinteq}), the reservoir solution $X_\omega(t)$ is most easily found by substituting $q(t)$ into (\ref{Xgensol}).

In deriving the frequency-domain result (\ref{qinteq}) we have indifferently commuted time and frequency integrations; if this commutation is not valid then the solution $q(\omega')$ of (\ref{qinteq}) will not give solutions $q(t)$ and $X_\omega(t)$ of the dynamical equations (\ref{qeq}) and (\ref{Xeq}). It turns out that for a constant (i.e.\ frequency-independent) coupling $\alpha(\omega)=a$, the time and frequency integrations do \emph{not} commute, but this case of constant coupling is easily solved in the time domain.  For (\ref{qinteq}) to be valid the coupling function must also be such that the integral containing $\alpha^2(\omega)$ converges. Needless to say, it must always be directly verified, regardless of the solution method employed, that the dynamical equations (\ref{qeq}) and (\ref{Xeq}) are satisfied.

Once a coupling function $\alpha(\omega)$ and the $t=0$ state of the reservoir are chosen, (\ref{qinteq}) presents a principal-value integral equation for $q(\omega')$, also known as a singular integral equation. In the standard classification, (\ref{qinteq}) is an inhomogeneous singular integral equation of the third kind. There is a systematic and very elegant method of solving singular integral equations, which exploits some complex analysis of Riemann and Hilbert~\cite{pipkin,muskhelishvili}. Integral equations such as (\ref{qinteq}) are solved by relating the equation to a boundary problem of the theory of analytic functions, usually called a Riemann problem but sometimes called a Hilbert problem~\cite{pipkin,muskhelishvili}. If the integral equation has a solution it can be found using any solution of the associated Riemann problem~\cite{pipkin,muskhelishvili}. We refer the reader to Pipkin's text~\cite{pipkin} for the details, which are too lengthy to be described here (the classic text of Muskhelishvili~\cite{muskhelishvili} does not appear to treat directly equations on an infinite interval). 

 The following additional remarks on the dynamical equations are worthwhile. Note that for any coupling function for which  (\ref{qinteq}) is valid, we can insert any function $q(\omega')(=q^*(-\omega'))$ and solve for $A_0(\omega')$ and $B_0(\omega')$, provided the integral involving $q(\omega')$ converges. In other words, we can \emph{choose} the dynamics of the $q$-oscillator from an enormous class of functions $q(t)$ whose Fourier transforms exist, and use (\ref{qinteq}) to find the values of the displacements and velocities of the reservoir at $t=0$ that will produce this dynamics $q(t)$. It must be borne in mind that there may well be solutions $q(t)$, $X_\omega(t)$ whose Fourier transforms do not exist; for example, there is no mathematical reason why the function $q(t)$ should be bounded. Even if $q(\omega)$ exists for a solution $q(t)$, it can only be found by solving (\ref{qinteq}) if the integral containing $q(\omega)$ is well defined. Transformation to the frequency domain is an essential part of diagonalizing the Hamiltonian, which in turn gives the most convenient description of quantum states of the system. But there is no guarantee that the diagonalized Hamiltonian will give all the dynamical solutions of the original Hamiltonian, and moreover there are couplings for which the Hamiltonian cannot be diagonalized (see section~\ref{sec:quantization}). 

It is already clear that our dynamical system is completely different from that in which  the continuum reservoir is replaced by a discrete one. The analysis of dynamical systems consisting of a discrete number of coupled harmonic oscillators is largely a matter of algebra~\cite{tat87}. In our case the mathematics is more challenging and this is because the dynamics is much richer. The possible couplings $\alpha(\omega)$ constitute the entire space of real functions, rather than a discrete set of numbers. Moreover, for some coupling functions $\alpha(\omega)$, the existence of a solution of (\ref{qinteq}) requires severe constraints on the functions $A_0(\omega')$ and $B_0(\omega')$, whereas for other coupling functions there will be no such constraints~\cite{pipkin}. The general mathematical theory also shows that there is a subspace of coupling functions for which (\ref{qinteq}) gives only the trivial solution $q(t)=X_\omega(t)=0$ when $X_\omega(0)=\dot{X}_\omega(0)=0$~\cite{pipkin}, although it is by no means clear what this subspace is. It would be very interesting to know how the space of solutions behaves as a functional of coupling $\alpha(\omega)$, but that is far beyond the scope of this paper. We will however find a constraint on the coupling function that must be satisfied for the Hamiltonian to be diagonalizable.

In addition to the form (\ref{Xgensol}), we also require the general solution of the reservoir equation (\ref{Xeq}) in a form that allows the imposition of conditions in the infinite past or future. We choose the infinite past $t\to -\infty$, and the required form of the general solution of (\ref{Xeq}) is obtained by using the retarded Green function (\ref{Gr}), yielding
\begin{equation}   \label{Xgensolret}
\fl
X_\omega(t)=A_{R}(\omega)\cos\omega t+B_{R}(\omega)\sin\omega t+\frac{\alpha(\omega)}{\omega}\int_{-\infty}^t\rmd t' \, q(t') \sin\left[\omega(t-t')\right]\,e^{-0^+(t-t')}.  
\end{equation}
Here the functions $A_{R}(\omega)$ and $B_{R}(\omega)$ can be viewed as determining $X_\omega(t)$  in the limit $t\to-\infty$, provided the last term vanishes as $t\to-\infty$ (this last property, however, may not hold for all solutions). The form (\ref{Xgensolret}) of the solution for the reservoir gives, in place of (\ref{qeqeff}), the $q$-oscillator equation
\begin{eqnarray}
\fl
\ddot{q}+\omega_0^2 q-\int_0^\infty\rmd\omega\int_{-\infty}^t\rmd t'\,q(t')\frac{\alpha^2(\omega)}{\omega} \sin\left[\omega(t-t')\right]\,e^{-0^+|t-t'|} \nonumber \\
-\int_0^\infty\rmd\omega\,\alpha(\omega)\left\{\frac{1}{2}\left[A_R(\omega)+\rmi B_R(\omega)\right]\exp(-\rmi\omega t)+\mathrm{c.c.}\right\}=0.  \label{qeqeffret}
\end{eqnarray}
In the frequency domain, (\ref{Xgensolret}) is
\begin{eqnarray}
\fl   X_\omega(\omega')=\pi\left[A_{R}(\omega)+\rmi B_{R}(\omega)\right]\delta(\omega-\omega')+\pi\left[A_{R}(\omega)-\rmi B_{R}(\omega)\right]\delta(\omega+\omega')
 \nonumber \\
+\mathrm{P}\frac{\alpha(\omega)q(\omega')}{\omega^2-{\omega'}^2}+ \frac{\rmi\pi \alpha(\omega)q(\omega')}{2\omega}\left[\delta(\omega-\omega')-\delta(\omega+\omega')\right],       \label{Xfreqgensolret}
\end{eqnarray}
which inserted into (\ref{qeqfreq}) gives the following equation for $q(\omega')$:
\begin{eqnarray}
\fl
\left[{\omega'}^2-\omega_0^2+\mathrm{P}\int_{0}^{\infty}\rmd \omega\frac{\alpha^2(\omega)}{\omega^2-{\omega'}^2}+\frac{\rmi\pi \alpha^2(|\omega'|)}{2\omega'}\right]q(\omega')  \nonumber \\
=-\pi\alpha(|\omega'|)\left[A_{R}(|\omega'|)+\rmi \,\mathrm{sgn}(\omega')B_{R}(|\omega'|)\right].  \label{qfreqret}
\end{eqnarray}
This is the frequency-domain version of (\ref{qeqeffret}). The solution $q(\omega')$ of (\ref{qfreqret}) gives the reservoir solution $X_\omega(t)$ by substitution of $q(t)$ into (\ref{Xgensolret}). This second formulation of the dynamical equations presents a much easier path to solutions, as we avoid the integral equation (\ref{qinteq}). Moreover, as every solution can in principle be obtained by solving either (\ref{qinteq}) or (\ref{qfreqret}), it may seem that (\ref{qinteq}) should be avoided entirely. But the ability in (\ref{qinteq}) to  impose directly conditions on the reservoir at a finite time allows the discovery of interesting particular solutions that in practice would not be found from (\ref{qfreqret}). This will be clearly demonstrated in sections~\ref{sec:solnI} and~\ref{sec:solnII}.

Before proceeding to quantize the system for a general coupling function $\alpha(\omega)$, we show classically that a particular coupling function gives  damping proportional to velocity. This coupling function will therefore allow an exact quantization of the dynamics (\ref{qdampf}).

\section{Coupling for damping proportional to velocity I}  \label{sec:solnI}
We consider the coupling function
\begin{equation}  \label{alpar}
\alpha(\omega)=\omega_0\omega\left[\frac{2\gamma}{\pi(\omega^2+\gamma^2)}\right]^{1/2}, \qquad \gamma>0,
\end{equation}
which increases monotonically from $0$ at $\omega=0$ to an asymptotic value of $\omega_0\sqrt{2\gamma/\pi}$ as $\omega\to\infty$. The function (\ref{alpar}) satisfies the integral relation
\begin{equation}  \label{Palpar}
\mathrm{P}\int_{0}^{\infty}\rmd \xi\frac{\alpha^2(\xi)}{\xi^2-{\omega}^2}=\frac{\gamma^2\omega_0^2}{\omega^2+\gamma^2},
\end{equation}
and for this choice of coupling the integral equation (\ref{qinteq}) reduces to
\begin{eqnarray}
\fl
\mathrm{P}\int_{-\infty}^{\infty}\rmd \xi\frac{q(\xi)}{\xi-\omega}=&-\frac{\pi\omega}{\gamma\omega_0^2}\left[\omega^2+\gamma^2-\omega_0^2\right]q(\omega)  \nonumber \\
&-\frac{\pi}{\omega_0}\left[\frac{2\pi}{\gamma}(\omega^2+\gamma^2)\right]^{1/2}\left[\mathrm{sgn}(\omega)A_0(|\omega|)+\rmi \,B_0(|\omega|)\right].  \label{qinteqpar}
\end{eqnarray}
The general solution of (\ref{qinteqpar}) will contain the solution of the homogeneous equation
\begin{equation}  \label{qinteqparhom}
\mathrm{P}\int_{-\infty}^{\infty}\rmd \xi\frac{q(\xi)}{\xi-\omega}=-\frac{\pi\omega}{\gamma\omega_0^2}\left[\omega^2+\gamma^2-\omega_0^2\right]q(\omega).
\end{equation}
We first solve the homogeneous equation (\ref{qinteqparhom}) and then solve the inhomogeneous equation (\ref{qinteqpar}) for arbitrary $A_0(\omega)$ and $B_0(\omega)$ by solving an associated integral equation for a type of Green function.

Following the standard procedure~\cite{pipkin}, the general solution of the homogeneous equation (\ref{qinteqparhom}) is found to be
\begin{equation} \label{qfreqparhom}
q(\omega)=\frac{a}{(\omega^2-\omega_0^2)^2+\gamma^2\omega^2},
\end{equation}
where $a$ is an arbitrary constant. Transforming (\ref{qfreqparhom}) to the time domain, we obtain three different behaviours depending on the size of the constant $\gamma>0$:
\numparts
\begin{numcases}{\!\!\!\!\!\!\!\!\!\!\!\!\!\! q(t)=}
b\exp\left(-\frac{\gamma |t|}{2}\right)\left[\cos\left(\frac{\omega_1 t}{2}\right)+\frac{\gamma}{\omega_1}\sin\left(\frac{\omega_1 |t|}{2}\right)\right],    &  for $\gamma < 2\omega_0$, \label{qtparhom1} \\
b\exp\left(-\omega_0 |t|\right)\left(1+\omega_0|t|\right), & for $\gamma = 2\omega_0$,  \label{qtparhom2} \\
c\exp\left(-\frac{\gamma |t|}{2}\right)\left[\exp\left(\frac{\gamma_1 |t|}{2}\right)-\frac{\gamma-\gamma_1}{\gamma+\gamma_1}\exp\left(-\frac{\gamma_1 |t|}{2}\right)\right], \!\!\!  &for $\gamma > 2\omega_0$, \label{qtparhom3}
\end{numcases}
\endnumparts
where $b$ and $c$ are arbitrary constants, and $\omega_1$ and $\gamma_1$ are defined by 
\begin{equation}  \label{omega1}
\omega_1=\sqrt{4\omega_0^2-\gamma^2}, \qquad \gamma_1=\sqrt{\gamma^2-4\omega_0^2}.
\end{equation}
Note that  $\omega_1$ in (\ref{qtparhom1}) is real, as is $\gamma_1$ in (\ref{qtparhom3}). The presence of the absolute value of $t$ in the solution (\ref{qtparhom1})--(\ref{qtparhom3}) may at a glance appear to give a discontinuity in $\dot{q}(t)$ and a delta function in $\ddot{q}(t)$, but in fact $\dot{q}(t)$ and $\ddot{q}(t)$ are continuous for all $t$.

The $q$-oscillator dynamics (\ref{qtparhom1})--(\ref{qtparhom3}) is exactly the solution of
\begin{equation}   \label{qtparhomeff}
\eqalign{ \ddot{q}+\gamma \dot{q}+\omega_0^2 q=0, \qquad t\ge 0,   \cr
\ddot{q}-\gamma \dot{q}+\omega_0^2 q=0, \qquad t\le 0, }
\end{equation}
with the condition $\dot{q}(0)=0$. We have thus obtained Ohmic damping and amplification of the  $q$-oscillator in a closed system that can be canonically  quantized. Recall that the time $t=0$, which divides the damping and amplification epochs in (\ref{qtparhomeff}), was an arbitrary choice. If we wish to impose arbitrary values of $q$ and $\dot{q}$ at some time $t_1$, with Ohmic damping for all later times, we need only replace $t=0$ in our derivations by a time $t_0\le t_1$, such that $t_0$ and the arbitrary constant in $q(t)$ give the required values of $q$ and $\dot{q}$ at $t_1$.

The dynamics (\ref{qtparhom1})--(\ref{qtparhom3}) is the complete solution for the $q$-oscillator when (\ref{qinteqpar}) is valid and $A_0(\omega)=B_0(\omega)=0$. This last condition sets the displacements and velocities of all the reservoir oscillators equal to zero at $t=0$ (see (\ref{Xgensol})). The solution for the reservoir is obtained from (\ref{Xgensol}) by substituting  (\ref{qtparhom1})--(\ref{qtparhom3}) and $A_0(\omega)=B_0(\omega)=0$; the result is
\numparts
\begin{eqnarray}
\fl
\gamma < 2\omega_0:  \nonumber \\
\fl
X_\omega(t)=\frac{b\omega\omega_0}{(\omega^2-\omega_0^2)^2+\gamma^2\omega^2}\left[\frac{2\gamma}{\pi(\omega^2+\gamma^2)}\right]^{1/2}\Bigg\{(\omega_0^2-\gamma^2-\omega^2)\cos (\omega t)+\frac{\gamma\omega_0^2}{\omega}\sin(\omega|t|)  \nonumber \\
\fl
\ \ \ +\left. \exp\left(-\frac{\gamma|t|}{2}\right)\left[(\omega^2+\gamma^2-\omega_0^2)\cos\left(\frac{\omega_1t}{2}\right)+\frac{\gamma}{\omega_1}(\omega^2+\gamma^2-3\omega_0^2)\sin\left(\frac{\omega_1|t|}{2}\right)\right]\right\} \label{Xtparhom1}  \\
\fl
\gamma = 2\omega_0:  \nonumber \\
\fl
X_\omega(t)=\frac{2b\omega\omega_0}{(\omega^2+\omega_0^2)^2}\left[\frac{\omega_0}{\pi(\omega^2+4\omega_0^2)}\right]^{1/2}\Bigg\{ \exp(-\omega_0|t|)\left[\omega^2+3\omega_0^2+(\omega^2+  
\omega_0^2)\omega_0|t|\right]    \nonumber \\
\left.-(\omega^2+3\omega_0^2)\cos(\omega t)+2\frac{\omega_0^3}{\omega}\sin(\omega |t|)\right\}  \label{Xtparhom2}   \\
\fl
\gamma > 2\omega_0:  \nonumber \\
\fl
X_\omega(t)=\frac{c\omega\omega_0}{(\omega^2-\omega_0^2)^2+\gamma^2\omega^2}\left[\frac{\gamma}{8\pi(\omega^2+\gamma^2)}\right]^{1/2}  \nonumber  \\
\fl
\quad \times\left\{\frac{8\gamma_1}{\gamma+\gamma_1}\left[(\omega_0^2-\gamma^2-\omega^2)\cos (\omega t)+\frac{\gamma\omega_0^2}{\omega}\sin(\omega|t|)\right]+ \exp\left(-\frac{\gamma|t|}{2}\right)\right.  \nonumber \\
\fl
\quad \left. \times\left[(4\omega^2+(\gamma+\gamma_1)^2)\exp\left(\frac{\gamma_1|t|}{2}\right)-\frac{\gamma-\gamma_1}{\gamma+\gamma_1}(4\omega^2+(\gamma-\gamma_1)^2)\exp\left(-\frac{\gamma_1|t|}{2}\right)\right]\right\}.  \label{Xtparhom3} 
\end{eqnarray}
\endnumparts

\begin{figure}[!htbp]
\begin{center} 
\includegraphics[width=9cm]{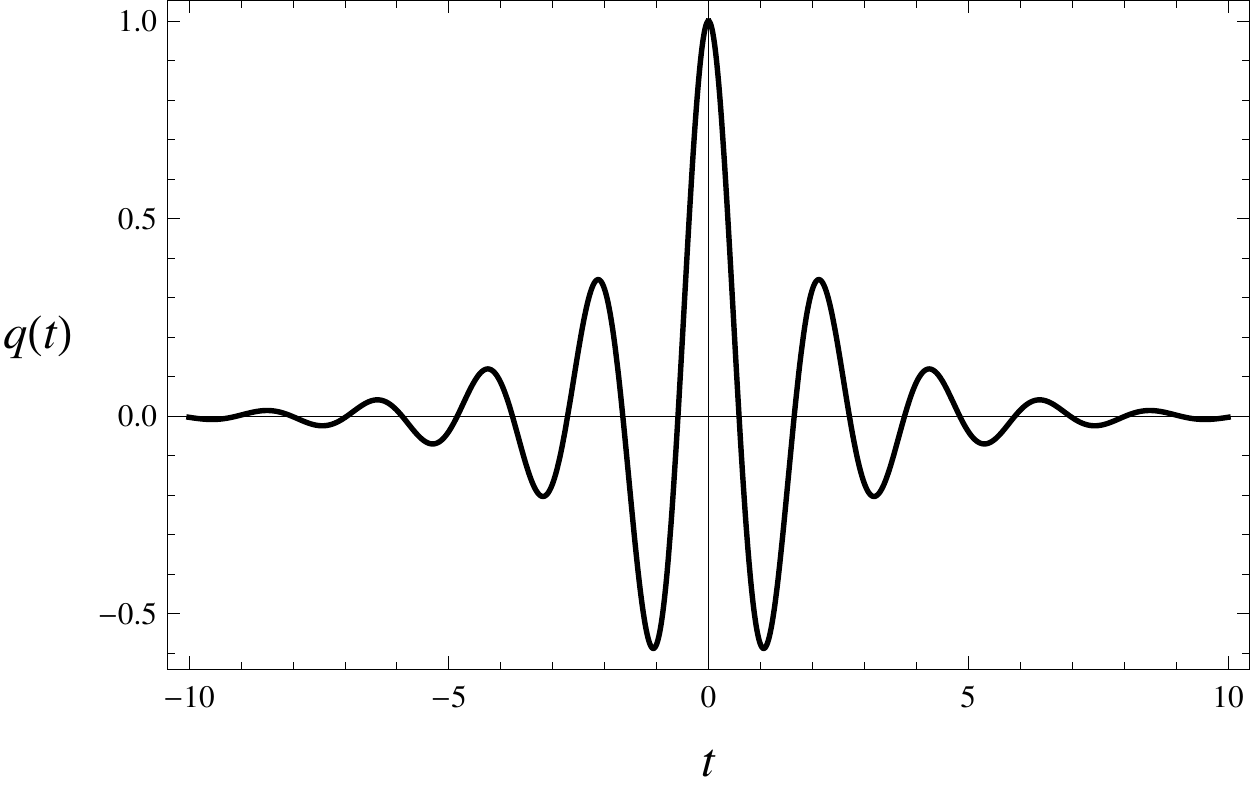}

\includegraphics[width=12cm]{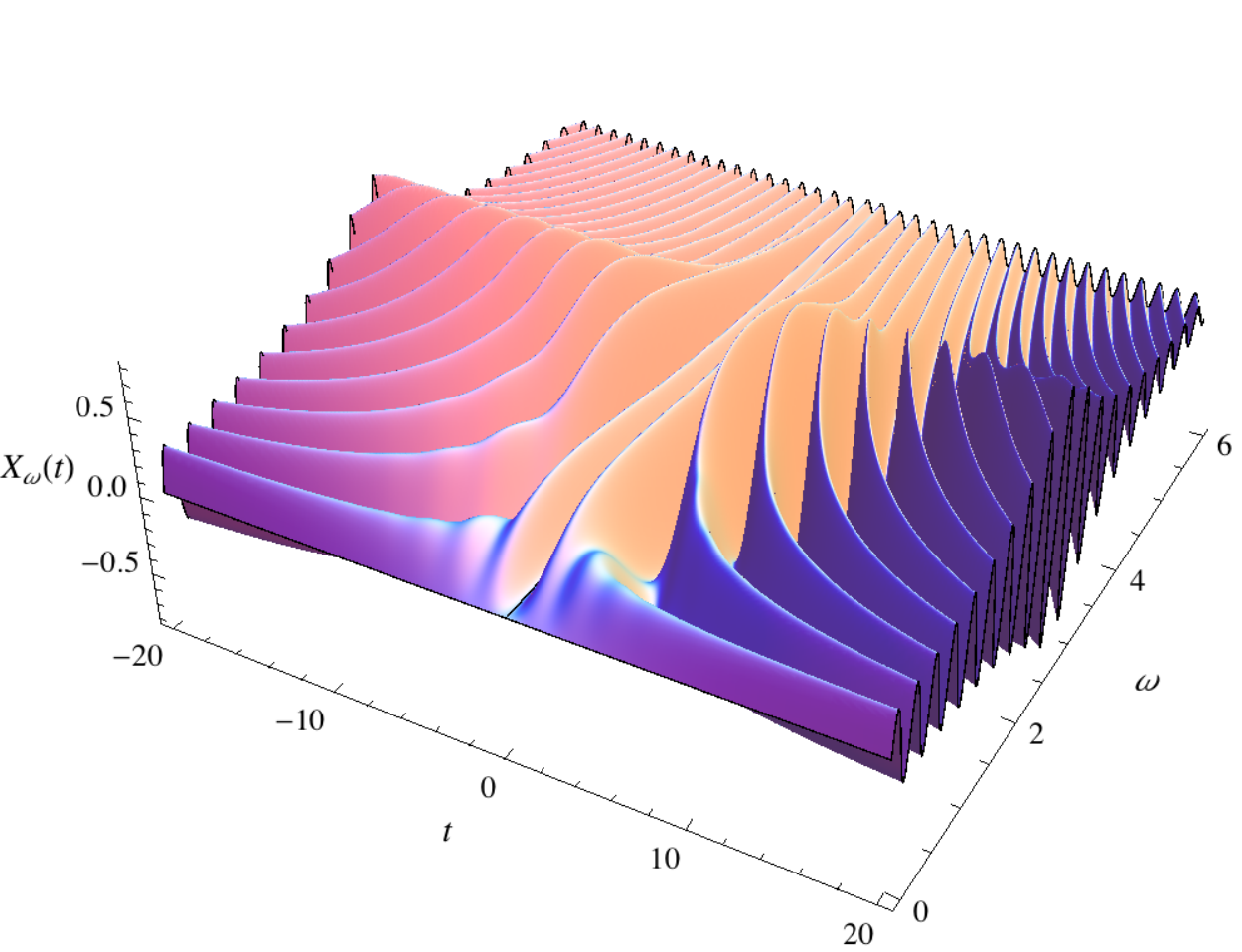}
\caption{The solution  (\ref{qtparhom1}) and (\ref{Xtparhom1}) with $\omega_0=3$, $\gamma=1$ and $b=1$. This solution is for coupling (\ref{alpar}) with $X_\omega(0)=\dot{X}_\omega(0)=0$. The $q$-oscillator is damped into the past and future from $t=0$, with damping proportional to velocity as shown in (\ref{qtparhomeff}). As $|t|\to\infty$, most of the $q$-oscillator energy goes into reservoir oscillators with frequencies around $\omega_0(=3)$.}
\label{fig:parsol}
\end{center}
\end{figure}

The under-damped case (\ref{qtparhom1}) and (\ref{Xtparhom1}) is plotted in Figure~\ref{fig:parsol} for a choice of parameters. Viewed from the infinite past to the infinite future, the $q$-oscillator is initially ($t\to-\infty$) at rest with all of the energy in the reservoir; the  $q$-oscillator is then amplified by the reservoir until at $t=0$ it has extracted all of the reservoir energy; then the energy is returned to the reservoir as the $q$-oscillator is damped. Alternatively, viewed as a consequence of the imposition of a $t=0$ condition, the $q$-oscillator is given a displacement, with $\dot{q}(0)=0$ and $X_\omega(0)=\dot{X}_\omega(0)=0$, and is then damped into the past and future, transferring its energy to the reservoir. Note in Figure~\ref{fig:parsol} that as the reservoir is amplified ($|t|$ increasing from $0$) and removes energy from the $q$-oscillator, the reservoir oscillators that attain the largest amplitudes as $|t|\to\infty$  are those with frequencies close to $\omega_0$, the free oscillation frequency of the $q$-oscillator. Compared to Figure~\ref{fig:parsol}, the over-damped case $\gamma \ge 2\omega_0$ has the following qualitative differences: the $q$-oscillator is exponentially damped into the past and future from $t=0$ without oscillating, and the amplitudes of the reservoir oscillators for $|t|\to\infty$ decrease as $\omega$ increases from $0$ (so there is no peak in the reservoir amplitudes at $\omega=\omega_0$).

The homogeneous version of equation (\ref{qeqeff}) (i.e.\ with $X_\omega(0)=\dot{X}_\omega(0)=0$) has in fact a slightly more general solution $q(t)$ than (\ref{qtparhom1})--(\ref{qtparhom3}) for coupling (\ref{alpar}). We did not obtain this more general solution in the frequency domain by solving the homogeneous integral equation (\ref{qinteqparhom}), because its Fourier transform $q(\omega)$ is not well enough behaved to make the derivation of (\ref{qinteqparhom}) valid. For coupling (\ref{alpar}), the time-domain equation (\ref{qeqeff}) is
\begin{equation}   \label{qeqeffpar}
\fl
\eqalign{  \ddot{q}+\omega_0^2 q-\mathrm{sgn}(t)\gamma\omega_0^2\int_0^t\rmd t'\,q(t')\exp(-\gamma|t-t'|)=f(t), \cr
f(t)=\int_0^\infty\rmd\omega\,\omega_0\omega\left[\frac{\gamma}{2\pi(\omega^2+\gamma^2)}\right]^{1/2}\left\{\left[A_0(\omega)+\rmi B_0(\omega)\right]\exp(-\rmi\omega t)+\mathrm{c.c.}\right\},   }
\end{equation}
which is (\ref{qinteqpar}) in the time domain. The homogeneous version of (\ref{qeqeffpar}) ($f(t)=0$) has a solution (\ref{qtparhom1})--(\ref{qtparhom3}), but the more general solution is 
\numparts
\begin{eqnarray} 
\fl
q(t)=&\frac{1}{2}(b_2-b_1)  \nonumber \\
\fl
&+\exp\left(-\frac{\gamma t}{2}\right)\left[b_1\cos\left(\frac{\omega_1 t}{2}\right)+\frac{b_1\gamma^2+(b_2-b_1)\omega_0^2}{\gamma\omega_1}\sin\left(\frac{\omega_1 t}{2}\right)\right], \qquad t\ge 0,   \label{qtparhomgen1}  \\
\fl
q(t)=&-\frac{1}{2}(b_2-b_1)     \nonumber \\
\fl
&+\exp\left(\frac{\gamma t}{2}\right)\left[b_2\cos\left(\frac{\omega_1 t}{2}\right)-\frac{b_2\gamma^2-(b_2-b_1)\omega_0^2}{\gamma\omega_1}\sin\left(\frac{\omega_1 t}{2}\right)\right], \qquad t\le 0,  \label{qtparhomgen2} 
\end{eqnarray}
\endnumparts
where $b_1$ and $b_2$ are arbitrary constants and $\omega_1$ is again given by (\ref{omega1}). For simplicity we have written (\ref{qtparhomgen1})--(\ref{qtparhomgen2}) in the form appropriate for the underdamped case $\gamma<2\omega_0$; this solution is also valid for $\gamma>2\omega_0$ and the solution for $\gamma=2\omega_0$ can be obtained by taking the limit $\omega_1\to 0$. The solution (\ref{qtparhom1})--(\ref{qtparhom3}) corresponds to the choice $b_2=b_1$ in (\ref{qtparhomgen1})--(\ref{qtparhomgen2}). The more general solution (\ref{qtparhomgen1})--(\ref{qtparhomgen2}) differs from (\ref{qtparhom1})--(\ref{qtparhom3}) by not having $\dot{q}(0)=0$ and by having a constant displacement of the $q$-oscillator as $|t|\to\infty$. Note that the displacement $q(t\to\infty)$ in (\ref{qtparhomgen1}) is minus the displacement $q(t\to-\infty)$ in (\ref{qtparhomgen2}). In the solution (\ref{qtparhomgen1})--(\ref{qtparhomgen2})  the reservoir extracts all the kinetic energy from the $q$-oscillator as $|t|\to\infty$ but does not bring it to its uncoupled equilibrium displacement $q=0$. By inserting  (\ref{qtparhomgen1})--(\ref{qtparhomgen2}) in  (\ref{Xgensol}) with $A_0(\omega)=B_0(\omega)=0$, the more general version of the solution (\ref{Xtparhom1})--(\ref{Xtparhom3}) for $X_\omega(t)$ is obtained, but we refrain from writing the result.

To complete the solution for the dynamics with coupling (\ref{alpar}), we must solve the equations of motion for non-zero $X_\omega(0)$ and $\dot{X}_\omega(0)$. We can write the solution in closed form for any $A_0(\omega)$ and $B_0(\omega)$ by finding a Green function for (\ref{qeqeffpar}). If we find a solution $G(t,t_0)$ of
\begin{equation}  \label{Gpareq}
\fl
\ddot{G}(t,t_0)+\omega_0^2 G(t,t_0)-\mathrm{sgn}(t)\gamma\omega_0^2\int_0^t\rmd t'\,G(t',t_0)\exp(-\gamma|t-t'|)=\delta(t-t_0),
\end{equation}
then the general solution of (\ref{qeqeffpar}) is
\begin{equation}
q(t)=\int_{-\infty}^\infty\rmd t_0\,G(t,t_0)f(t_0)
\end{equation}
plus the solution (\ref{qtparhomgen1})--(\ref{qtparhomgen2}) of the homogeneous version of (\ref{qeqeffpar}) (with $f(t)=0$). The solution for $X_\omega(t)$ is then obtained from (\ref{Xgensol}). A Green function $G(t,t_0)$ can be found by Fourier transforming (\ref{Gpareq}) in $t$ and solving the resulting integral equation for $G(\omega,t_0)$. As (\ref{qinteqpar}) is the Fourier transform in $t$ of (\ref{qeqeffpar}), we see from (\ref{Gpareq}) that the integral equation for $G(\omega,t_0)$ is
\begin{equation}
\fl
\mathrm{P}\int_{-\infty}^{\infty}\rmd \xi\frac{G(\xi,t_0)}{\xi-\omega}=-\frac{\pi\omega}{\gamma\omega_0^2}\left[\omega^2+\gamma^2-\omega_0^2\right]G(\omega,t_0)    -\frac{\pi(\omega^2+\gamma^2)}{\gamma\omega_0^2(\omega-\rmi\epsilon)}\exp(\rmi\omega t_0),  \label{Ginteq}
\end{equation}
where a pole at $\omega=0$ in the last term has been moved off the real line by inserting $\epsilon\neq0$. The two choices of sign of $\epsilon$ in (\ref{Ginteq}) give two Green functions $G(t,t_0)$ in the time domain that differ by a solution of the homogeneous version of (\ref{Gpareq}) (with 0 on the right-hand side); this homogeneous solution is of course given by (\ref{qtparhomgen1})--(\ref{qtparhomgen2}) with a particular $b_1$ and $b_2$. The Green functions have rather lengthy expressions and the results for $G(\omega,t_0)$ and $G(t,t_0)$ are given in \ref{ap:green} for the choice $\epsilon>0$. 

Although the classical dynamics is now completely solved for the coupling (\ref{alpar}), the diagonalization of the Hamiltonian (in both the classical and quantum cases) will make use of an alternative formulation based on (\ref{Xgensolret}).

\section{Coupling for damping proportional to velocity II}  \label{sec:solnII}
The forms (\ref{Xgensolret}) and (\ref{qeqeffret}) of the dynamical equations can be obtained by pushing back the (arbitrary) integration limit $t=0$ in (\ref{Xgensol}) and (\ref{qeqeff}) to $t\to-\infty$. The part of $X_\omega(t)$ in (\ref{Xgensolret}) that is the solution of the homogeneous version of (\ref{Xeq}) (with $q(t)=0$) describes the state of the reservoir at $t\to-\infty$ if the last term in  (\ref{Xgensolret}) vanishes as $t\to-\infty$. 

For the coupling (\ref{alpar}), which satisfies (\ref{Palpar}), the $q$-oscillator equation (\ref{qfreqret}) reduces to
\begin{equation}   \label{qfreqretpar}
\eqalign{ \frac{\omega}{\omega+\rmi\gamma}(\omega^2+\rmi\gamma\omega-\omega_0^2)q(\omega) =g(\omega), \cr
g(\omega)=-\omega\omega_0\left[\frac{2\pi\gamma}{\omega^2+\gamma^2}\right]^{1/2}\left[\mathrm{sgn}(\omega) A_R(|\omega|)+\rmi B_R(|\omega|)\right].   }
\end{equation}
In the time domain this is (\ref{qeqeffret}) with coupling (\ref{alpar}):
\begin{eqnarray}  
\fl
  \ddot{q}+\omega_0^2 q-\gamma\omega_0^2\int_{-\infty}^t\rmd t'\,q(t')\exp[-\gamma(t-t')]=g(t),   \label{qeqeffretpar}   \\
  \fl
g(t)=\int_{0}^\infty\rmd\omega\,\omega_0\omega\left[\frac{\gamma}{2\pi(\omega^2+\gamma^2)}\right]^{1/2}\left\{\left[A_R(\omega)+\rmi B_R(\omega)\right]\exp(-\rmi\omega t)+\mathrm{c.c.}\right\}.     \label{gtdef} 
\end{eqnarray}

Because of the factor of $\omega$ on the left-hand side of (\ref{qfreqretpar}), the homogeneous equation ($g(\omega)=0$) has the solution
\begin{equation}   \label{qzm}
q(\omega)=2\pi a\delta(\omega) \qquad \Longrightarrow \qquad q(t)=a,
\end{equation}
where $a$ is an arbitrary constant. The corresponding solution for the reservoir is obtained from (\ref{Xgensolret}) with $q(t)=a$ and $A_R(\omega)=B_R(\omega)=0$; the reservoir is also independent of time and the full solution is
\begin{equation}  \label{zm}
q(t)=a,  \qquad X_\omega(t)=\frac{a\omega_0}{\omega}\left[\frac{2\gamma}{\pi(\omega^2+\gamma^2)}\right]^{1/2}.
\end{equation}
This ``zero-mode" solution has zero energy, as can be verified using (\ref{E}). It is interesting to check when a zero mode occurs in the general dynamical equations (\ref{qeq}) and (\ref{Xeq}). Inserting $q(t)=a$ into (\ref{qeq}) and (\ref{Xeq}) one finds that for $a\neq 0$ there exists a consistent solution
\begin{equation}
q(t)=a,  \qquad X_\omega(t)=a\frac{\alpha(\omega)}{\omega^2}
\end{equation}
if and only if the coupling function satisfies
\begin{equation}  \label{alcon}
\omega_0^2=\int_0^\infty\rmd \omega \frac{\alpha^2(\omega)}{\omega^2}.
\end{equation}
The coupling (\ref{alpar}) is one example of a function that satisfies (\ref{alcon}), and this is why there is a zero mode (\ref{zm}). Zero modes are always a possibility for oscillators coupled as in (\ref{L}); even for just two oscillators coupled in this fashion there exists a zero mode for one value of the square of the coupling constant. In the  case of the system (\ref{L}) there is an infinite class of coupling functions that give a zero mode, and an infinite class that do not. It is interesting that the coupling function that gives Ohmic damping of the $q$-oscillator also allows a zero mode, but the wider significance of this fact is not immediately clear. Note that the zero mode (\ref{zm}) is not related to the constant displacement as $t\to\pm\infty$ of the $q$-oscillator in the solution  (\ref{qtparhomgen1})--(\ref{qtparhomgen2}); that solution has $X_\omega(0)=\dot{X}_\omega(0)=0$ and so has no relation to (\ref{zm}). The formulation of the dynamical equations used in this section has shown that the zero mode (\ref{zm}) is completely determined by the condition $A_R(\omega)=B_R(\omega)=0$ on the reservoir, and we see from (\ref{zm}) that this is not equivalent to the condition $X_\omega(t)|_{t\to-\infty}=0$.

The general solution of (\ref{qfreqretpar}) is the solution (\ref{qzm}) of the homogeneous equation plus the solution
\begin{eqnarray}
q(\omega)=-G_R(\omega)g(\omega),    \label{qfreqsolret} \\
G_R(\omega)=-\frac{\omega+\rmi\gamma}{(\omega+\rmi 0^+)(\omega^2+\rmi\gamma\omega-\omega_0^2)}.  \label{GRfreq}
\end{eqnarray}
In (\ref{GRfreq}) we have defined a Green function and given a prescription for dealing with its pole at $\omega=0$. We have moved the $\omega=0$ pole into the lower-half complex $\omega$-plane, where the other two poles of $G_R(\omega)$ lie, and have therefore chosen the retarded Green function (analytic in the upper-half $\omega$-plane). It is easy to see that a different prescription for dealing with the $\omega=0$ pole will give a Green function that differs from (\ref{GRfreq}) by a solution (\ref{qzm}) of the homogeneous equation. The Green function (\ref{GRfreq}) in the time domain is
\begin{equation}
\fl
G_R(t)=\theta(t)\frac{\gamma}{\omega_0^2}\left\{e^{-0^+t}- \exp\left(-\frac{\gamma t}{2}\right)\left[\cos\left(\frac{\omega_1 t}{2}\right)+\frac{\gamma^2-2\omega_0^2}{\gamma\omega_1}\sin\left(\frac{\omega_1 t}{2}\right)\right]  \right\},
  \label{GRt}
\end{equation}
with $\omega_1$ given by (\ref{omega1}). The result (\ref{GRt}) is valid for all values of $\gamma$ if the case $\gamma=2\omega_0$ is understood as the limit $\omega_1\to 0$. As is expected from (\ref{qeqeffretpar}), the Green function (\ref{GRt}) satisfies
\begin{equation}  \label{GReq}
 \ddot{G_R}+\omega_0^2 G_R-\gamma\omega_0^2\int_{-\infty}^t\rmd t'\,G_R(t')\exp[-\gamma(t-t')]=\delta(t).
\end{equation}
In the time domain the general solution for the  $q$-oscillator is
\begin{equation}  \label{qtsolret}
q(t)=a+\int_{-\infty}^\infty\rmd t'\,G_R(t-t')g(t'),
\end{equation}
as can be confirmed by showing that the second term satisfies (\ref{qeqeffretpar}) because of (\ref{GReq}). The solution for $X_\omega(t)$ is given by (\ref{Xgensolret}).

By removing the step-function factor $\theta(t)$ in the retarded Green function (\ref{GRt}), we obtain a solution of the homogeneous version of (\ref{GReq}) (with zero on the right-hand side).  This is of course also a solution of the homogeneous version of (\ref{qeqeffretpar}), so there seems to be a more general solution for $q(t)$  with $A_R(\omega)=B_R(\omega)=0$ than the zero-mode solution (\ref{qzm}). In fact the general solution of the homogeneous version of (\ref{qeqeffretpar}) is given by (\ref{qtparhomgen1}), with the expression valid for all $t$ rather than $t\geq 0$. (The Green function (\ref{GRt}) with $\theta(t)$ removed corresponds to (\ref{qtparhomgen1}) with $b_1=-\gamma/\omega_0^2$ and $b_2=\gamma/\omega_0^2$.) But the expression in (\ref{qtparhomgen1}), taken to be valid for all $t$, is not a solution for the dynamical system for the simple reason that the corresponding reservoir solution, given by  (\ref{Xgensolret}), diverges. This behaviour of the dynamics can be described with reference to Fig.~\ref{fig:parsol}, which shows a particular case of the solution (\ref{qtparhomgen1})--(\ref{qtparhomgen2}) and the corresponding reservoir solution. The time $t=0$ in Fig.~\ref{fig:parsol} is arbitrary and the solution holds with the $t=0$ peak in $q(t)$, and zero in $X_\omega(t)$, moved to any other finite value of $t$. If this peak in $q(t)$ is pushed back to $t\to-\infty$, however, the amplitude of the peak diverges and this causes the entire solution  $X_\omega(t)$ to diverge. The general solution for the dynamics with the condition $A_R(\omega)=B_R(\omega)=0$ is therefore the zero mode (\ref{zm}).

The solution shown in Fig.~\ref{fig:parsol} must be a particular case of the general solution as formulated in this section: it must correspond to some choice of $A_R(\omega)$ and $B_R(\omega)$. As the solution in Fig.~\ref{fig:parsol} has $q(t)\to 0$ as $|t|\to\infty$, the corresponding functions $A_R(\omega)$ and $B_R(\omega)$ turn out to specify the state of the reservoir at $t\to-\infty$ (see (\ref{Xgensolret})). By taking the limit $t\to-\infty$ in (\ref{Xtparhom1}) we see that $X_\omega(t)$ takes the form of the first two terms in (\ref{Xgensolret}), with
\begin{eqnarray}
A_R(\omega)=-b\left[\frac{2\gamma}{\pi(\omega^2+\gamma^2)}\right]^{1/2}\frac{\omega\omega_0(\omega^2+\gamma^2-\omega_0^2)}{(\omega^2-\omega_0^2)^2+\gamma^2\omega^2},  \label{ARpar} \\
B_R(\omega)=-b\left[\frac{2\gamma}{\pi(\omega^2+\gamma^2)}\right]^{1/2}\frac{\gamma\omega_0^3}{(\omega^2-\omega_0^2)^2+\gamma^2\omega^2}.   \label{BRpar}
\end{eqnarray}
One can verify that with $A_R(\omega)$ and $B_R(\omega)$ given by (\ref{ARpar}) and (\ref{BRpar}), and with $a=0$, equations (\ref{qtsolret}), (\ref{GRt}), (\ref{gtdef}) and (\ref{Xgensolret}) reproduce the solution (\ref{qtparhom1})--(\ref{qtparhom3}) and (\ref{Xtparhom1})--(\ref{Xtparhom3}). Needless to say, it is by far from obvious that the specification (\ref{ARpar}) and (\ref{BRpar}) of the reservoir at $t\to-\infty$ implies the interesting behaviour depicted in Fig.~\ref{fig:parsol} (modulo a zero-mode solution (\ref{zm})).

We have formulated the general solution for the dynamics with coupling (\ref{alpar}) in two ways. Every particular solution is completely specified by either the functions $A_0(\omega)$ and $B_0(\omega)$ and constants $b_1$ and $b_2$ in section~\ref{sec:solnI}, or by the functions $A_R(\omega)$ and $B_R(\omega)$ and constant $a$ in this section. The relationship between the pairs $\{A_0(\omega),B_0(\omega)\}$ and $\{A_R(\omega),B_R(\omega)\}$ for the \emph{same} solution $\{q(t),X_\omega(t)\}$ is the one that transforms (\ref{Xfreqgensol}) into (\ref{Xfreqgensolret}) and (\ref{qinteq}) into (\ref{qfreqret}); it is easy to show that this relationship is
\begin{eqnarray}
\fl
A_0(|\omega|)= A_R(|\omega|)-\frac{\alpha(|\omega|)}{2\omega}\left\{\mathrm{Im}[q(\omega)]+\frac{2\omega}{\pi}\mathrm{P}\int_0^\infty\rmd \xi \, \frac{\mathrm{Re}[q(\xi)]}{\xi^2-\omega^2}\right\},
  \label{A0AR} \\
  \fl
B_0(|\omega|)= B_R(|\omega|)+\mathrm{sgn}(\omega)\frac{\alpha(|\omega|)}{2\omega}\left\{\mathrm{Re}[q(\omega)]-\frac{2}{\pi}\mathrm{P}\int_0^\infty\rmd \xi \, \frac{\xi\,\mathrm{Im}[q(\xi)]}{\xi^2-\omega^2}\right\}.   \label{B0BR}
\end{eqnarray}
The results (\ref{ARpar}) and (\ref{BRpar}) for the particular solution (\ref{qtparhom1})--(\ref{qtparhom3}) and (\ref{Xtparhom1})--(\ref{Xtparhom3}) can also be derived from (\ref{A0AR}) and (\ref{B0BR}) by inserting $A_0(\omega)=B_0(\omega)=0$, (\ref{alpar}) and (\ref{qfreqparhom}).

\section{Quantization and diagonalization of the Hamiltonian}   \label{sec:quantization}
The canonical momenta for the Lagrangian (\ref{L}) are
\begin{equation}  \label{canmom}
\Pi_q(t)=\dot{q}(t), \qquad \Pi_{X_\omega}(t)=\dot{X_\omega}(t).
\end{equation}
We quantize the system in the Heisenberg picture by imposing the equal-time canonical commutation relations
\begin{eqnarray}
[\hat{q}(t),\hat{\Pi}_q(t)]=\rmi\hbar, \qquad [\hat{X}_\omega(t),\hat{\Pi}_{X_{\omega'}}(t)]=\rmi\hbar\,\delta(\omega-\omega'),  \label{cancom1} \\
\fl
{[}\hat{X}_\omega(t),\hat{X}_{\omega'}(t)]=0, \quad  [\hat{\Pi}_{X_{\omega}}(t),\hat{\Pi}_{X_{\omega'}}(t)]=0, \quad [\hat{q}(t),\hat{X}_\omega(t)]=0, \quad [\hat{q}(t),\hat{\Pi}_{X_{\omega}}(t)]=0.  \label{cancom2} 
\end{eqnarray}
The commutation relations for the reservoir are similar to that of a field theory because of the continuum of frequencies $\omega$. The Hamiltonian is
\begin{equation}   \label{H}
\fl
\hat{H}=\frac{1}{2}\hat{\Pi}_q^2+\frac{1}{2}\omega_0^2 \hat{q}^2+\frac{1}{2}\int_0^\infty\rmd\omega\left(\hat{\Pi}_{X_{\omega}}^2+\omega^2\hat{X}_\omega^2\right)-\frac{1}{2}\int_0^\infty\rmd\omega\,\alpha(\omega)\left[\hat{q}\hat{X}_\omega+\hat{X}_\omega\hat{q}\right],
\end{equation}
where a Hermitian combination of the operators has been taken in the last term; classically, (\ref{H}) gives the total energy (\ref{E}).

The solution of Hamilton's equations for the canonical operators can be immediately written down using the classical results in sections~\ref{sec:solnI} and~\ref{sec:solnII}. But in order to describe general quantum states we need some basis in the Hilbert space, and a solution for the canonical operators does not by itself reveal such a basis. Diagonalization of the Hamiltonian solves Hamilton's equations for the canonical operators in terms of the energy eigenstates of the system and thus solves the dynamics of general quantum states. The diagonalization process can also be performed on a purely classical level, where the classical normal modes of the system correspond to the quantum energy eigenstates. What is required is a new set of dynamical variables for which the system reduces to a set of uncoupled harmonic oscillators; each of these free oscillations (normal modes) has a conserved energy and gives an energy eigenstate in the quantum theory. 

For a discrete set of oscillators with coupling terms of the same form as in (\ref{L}), the normal-mode displacements are linear combinations of the displacements of the coupled oscillators~\cite{tat87}. The continuum reservoir in (\ref{L}), however, has the remarkable effect that the normal modes are given by a \emph{canonical} transformation of the original variables, in which displacements and canonical momenta are mixed. It is well known that in the discrete case the frequencies of the normal modes become complex when the coupling of the form (\ref{L}) is too large~\cite{tat87}. We shall also find that there is a restriction on the coupling function $\alpha(\omega)$ in order for the Hamiltonian (\ref{H}) to be diagonalizable with real-frequency eigenmodes. This restriction on $\alpha(\omega)$ is met by the particular coupling (\ref{alpar}). In contrast, Huttner and Barnett~\cite{hut92} showed, as part of their model of a dielectric, that the system (\ref{L}) with $X_\omega$ in the coupling term replaced by $\dot{X}_\omega$ has real-frequency eigenmodes for essentially any coupling function $\alpha(\omega)$.

We attempt to show that the Hamiltonian (\ref{H}) can be written
\begin{eqnarray}
\hat{H}& =\frac{1}{2}\int_0^\infty\rmd\omega\left(\hat{\Pi}_{\Phi_{\omega}}^2+\omega^2\hat{\Phi}_\omega^2\right)   \label{HPhi}  \\
&=\frac{1}{2}\int_0^\infty\rmd\omega\,\hbar\omega\left[\hat{C}^\dagger(\omega,t)\hat{C}(\omega,t)+\hat{C}(\omega,t)\hat{C}^\dagger(\omega,t)\right],   \label{HC} 
\end{eqnarray}
where $\hat{\Phi}_\omega(t)$, $\omega\in[0,\infty)$ are displacement operators for a set of uncoupled harmonic oscillators, the normal modes or energy eigenmodes of the system. The normal-mode creation and annihilation operators $\hat{C}^\dagger(\omega,t)$ and $\hat{C}(\omega,t)$ are given by the usual harmonic-oscillator expressions
\begin{equation}  \label{Cdef}
\fl
\hat{C}(\omega,t)=\sqrt{\frac{\omega}{2\hbar}}\left[\hat{\Phi}_\omega(t)+\frac{\rmi}{\omega}\hat{\Pi}_{\Phi_{\omega}}(t)\right], \qquad \hat{C}^\dagger(\omega,t)=\sqrt{\frac{\omega}{2\hbar}}\left[\hat{\Phi}_\omega(t)-\frac{\rmi}{\omega}\hat{\Pi}_{\Phi_{\omega}}(t)\right],
\end{equation}
and they obey the commutation relations
\begin{equation}   \label{Ccom} 
[\hat{C}(\omega,t),\hat{C}^\dagger(\omega',t)]=\delta(\omega-\omega'), \qquad  [\hat{C}(\omega,t),\hat{C}(\omega',t)]=0.
\end{equation}
From (\ref{HC}) and (\ref{Ccom}) we obtain
\begin{equation}  \label{CH}
[\hat{C}(\omega,t),\hat{H}]=\hbar\omega\hat{C}(\omega,t),
\end{equation}
so that the eigenmode creation and annihilation operators have a stationary time-dependence given by
\begin{equation}  \label{Ct}
\hat{C}(\omega,t)=\exp(-\rmi\omega t)\hat{C}(\omega).
\end{equation}
The diagonalized form (\ref{HPhi}) and (\ref{HC}) of the Hamiltonian (\ref{H}) will be achieved with a restriction on the coupling function $\alpha(\omega)$. The following derivation can be performed classically if commutators are replaced with Poisson brackets.

If the diagonalization is possible then there exists a transformation from the original dynamical degrees of freedom in (\ref{H}) to eigenmode degrees of freedom for which the Hamiltonian takes the form (\ref{HPhi}). In the quantum theory it is convenient to use the creation and annihilation operators (\ref{Cdef}) as the eigenmode degrees of freedom, rather than the canonical operators $\hat{\Phi}_\omega$ and $\hat{\Pi}_{\Phi_{\omega}}$ (the quantities (\ref{Cdef}) can also be used in the classical theory). We thus seek a transformation
\begin{eqnarray}
\fl
\hat{q}(t)=\int_0^\infty\rmd\omega\left[f_q(\omega)\hat{C}(\omega,t)+\mathrm{h.c.}\right],     \qquad  \hat{\Pi}_q(t)=\int_0^\infty\rmd\omega\left[f_{\Pi_q}(\omega)\hat{C}(\omega,t)+\mathrm{h.c.}\right],    \label{qfC1}  \\
\hat{X}_\omega(t)=\int_0^\infty\rmd\omega'\left[f_X(\omega,\omega')\hat{C}(\omega',t)+\mathrm{h.c.}\right],     \label{qfC2} \\
  \hat{\Pi}_{X_\omega}(t)=\int_0^\infty\rmd\omega'\left[f_{\Pi_X}(\omega,\omega')\hat{C}(\omega',t)+\mathrm{h.c.}\right],   \label{qfC3}
\end{eqnarray}
for which (\ref{H}) becomes (\ref{HC}). The unknown coefficients $f_q(\omega)$, etc., in (\ref{qfC1})--(\ref{qfC3}) can be written as commutators by utilizing (\ref{Ccom}):
\begin{eqnarray}
f_q(\omega)=[\hat{q}(t),\hat{C}^\dagger(\omega,t)],     \qquad  f_{\Pi_q}(\omega)=[\hat{\Pi}_q(t),\hat{C}^\dagger(\omega,t)],  \label{fC1} \\
f_X(\omega,\omega')=[\hat{X}_\omega(t),\hat{C}^\dagger(\omega',t)],     \qquad  f_{\Pi_X}(\omega,\omega')= [\hat{\Pi}_{X_\omega}(t),\hat{C}^\dagger(\omega',t)].  \label{fC2}
\end{eqnarray}
The eigenmodes variables must be expressible in terms of the original variables, i.e.\  (\ref{qfC1})--(\ref{qfC3}) must be invertible, and (\ref{fC1}), (\ref{fC2}), (\ref{cancom1}) and (\ref{cancom2}) imply
\begin{eqnarray}
\fl
\hat{C}(\omega,t)=-\frac{\rmi}{\hbar}{\Bigg\{} & f^*_{\Pi_q}(\omega)\hat{q}(t)-f^*_q(\omega)\hat{\Pi}_q(t)    \nonumber \\
\fl
&   \left.    +\int_0^\infty\rmd\omega'\left[f^*_{\Pi_X}(\omega',\omega)\hat{X}_{\omega'}(t)-f^*_X(\omega',\omega)\hat{\Pi}_{X_{\omega'}}(t)\right]\right\}.   \label{Cfq}
\end{eqnarray}

It is clear from (\ref{qfC1})--(\ref{qfC3}) and (\ref{Ct}) that the $f$-coefficients are closely related to Fourier transforms of the original canonical operators. This suggests that the  $f$-coefficients must satisfy the original dynamical equations written in the frequency domain. We derive these equations for the $f$-coefficients as follows. By inserting (\ref{Cfq}) and (\ref{H}) in (\ref{CH}), and using the canonical commutation relations (\ref{cancom1}) and (\ref{cancom2}), we obtain
\begin{eqnarray}
\fl
\hbar\omega\hat{C}(\omega,t)= f^*_{\Pi_q}(\omega)\hat{\Pi}_q(t)+\omega_0^2f^*_q(\omega)\hat{q}(t)  
    +\int_0^\infty\rmd\omega'\left\{f^*_{\Pi_X}(\omega',\omega)\hat{\Pi}_{X_{\omega'}}(t)   \right.     \nonumber \\
 \left.   +{\omega'}^2f^*_X(\omega',\omega)\hat{X}_{\omega'}(t)-\alpha(\omega')\left[f^*_q(\omega)\hat{X}_{\omega'}(t)+f^*_X(\omega',\omega)\hat{q}(t)\right]\right\}.   \label{Cfq2}
\end{eqnarray}
Comparing coefficients of the canonical operators in (\ref{Cfq}) and (\ref{Cfq2}) we find
\begin{eqnarray}
\fl
f_{\Pi_q}(\omega)=-\rmi\omega f_q(\omega), \qquad  \rmi\omega  f_{\Pi_q}(\omega)=\omega_0^2 f_q(\omega)- \int_0^\infty\rmd\omega' \alpha(\omega') f_X(\omega',\omega),  \label{fPifq} \\ 
\fl
f_{\Pi_X}(\omega',\omega)=-\rmi\omega f_X(\omega',\omega), \qquad  \rmi\omega f_{\Pi_X}(\omega',\omega)={\omega'}^2f_X(\omega',\omega)- \alpha(\omega') f_q(\omega),   \label{fPifX}
\end{eqnarray}
which give
\begin{eqnarray}
\omega^2 f_q(\omega)=\omega_0^2 f_q(\omega)- \int_0^\infty\rmd\omega' \alpha(\omega') f_X(\omega',\omega), \label{fqeq}  \\
{\omega}^2f_X(\omega',\omega)={\omega'}^2f_X(\omega',\omega)- \alpha(\omega') f_q(\omega).   \label{fXeq}
\end{eqnarray}
These two equations for $f_q(\omega)$ and $f_X(\omega',\omega)$ are indeed identical to the frequency-domain equations (\ref{qeqfreq}) and (\ref{Xeqfreq}) for  $q(\omega)$ and $X_{\omega'}(\omega)$. We can therefore write the general solution of  (\ref{fqeq}) and (\ref{fXeq}) using the results (\ref{Xfreqgensolret}) and (\ref{qfreqret}), in which a retarded Green function was used to solve the reservoir equation. As only positive frequency arguments are used in the $f$-coefficients, we can drop delta functions containing sums of frequencies and so (\ref{Xfreqgensolret}) gives the following general solution for $f_X(\omega',\omega)$:
\begin{eqnarray}
\fl
f_X(\omega',\omega)&=h_X(\omega)\delta(\omega-\omega')+\mathrm{P}\frac{\alpha(\omega') }{{\omega'}^2-\omega^2}f_q(\omega)+\frac{\rmi\pi \alpha(\omega')}{2\omega'}f_q(\omega)\delta(\omega-\omega')  \label{fXsol0} \\
\fl
&=h_X(\omega)\delta(\omega-\omega')+\frac{\alpha(\omega')}{2\omega'}\left(\frac{1}{\omega'-\omega-\rmi 0^+}+\frac{1}{\omega'+\omega}\right)f_q(\omega),   \label{fXsol}
\end{eqnarray}
where $h_X(\omega)$ is an arbitrary complex function. The corresponding general solution for $f_q(\omega)$ is found from (\ref{fqeq}) and (\ref{fXsol0}), which yields (\ref{qfreqret}) with positive frequency arguments:
\begin{equation}   \label{fqeffeq}
\left[{\omega}^2-\omega_0^2+\mathrm{P}\int_{0}^{\infty}\rmd \xi\frac{\alpha^2(\xi)}{\xi^2-{\omega}^2}+\frac{\rmi\pi \alpha^2(\omega)}{2\omega}\right]f_q(\omega)=-\alpha(\omega)h_X(\omega).
\end{equation}
The solution of (\ref{fqeffeq}) can be written
\begin{equation}   \label{fqsol}
\eqalign{ f_q(\omega)=h_q(\omega)+G(\omega)\alpha(\omega)h_X(\omega),  \cr
G(\omega)=-\left[{\omega}^2-\omega_0^2+\mathrm{P}\int_{0}^{\infty}\rmd \xi\frac{\alpha^2(\xi)}{\xi^2-{\omega}^2}+\frac{\rmi\pi \alpha^2(\omega)}{2\omega}\right]^{-1}, }
\end{equation}
where we have defined a Green function $G(\omega)$, and $h_q(\omega)$ is the solution of 
\begin{equation}   \label{hqeq}
\left[{\omega}^2-\omega_0^2+\mathrm{P}\int_{0}^{\infty}\rmd \xi\frac{\alpha^2(\xi)}{\xi^2-{\omega}^2}+\frac{\rmi\pi \alpha^2(\omega)}{2\omega}\right]h_q(\omega)=0.
\end{equation}
The existence of a non-zero solution of (\ref{hqeq}) requires at least one real root of the function in brackets; the resulting $h_q(\omega)$ will contain a delta-function factor and an example is the zero mode (\ref{qzm}) for the coupling (\ref{alpar}). The $f$-coefficients are now completely determined by the two functions $h_X(\omega)$ and $h_q(\omega)$; it remains to show that values for these two functions exist that give the required diagonalization. As follows from section~\ref{sec:L}, there are alternative ways of writing the general solution for the $f$-coefficients, but it turns out that the form (\ref{fXsol0})--(\ref{hqeq}), based on the retarded solution of the reservoir equation, is very convenient for solving the diagonalization problem. An equally convenient approach, which is in fact adopted by Huttner and Barnett~\cite{hut92}, is based on the advanced solution of the reservoir equation.

A value for the function $h_X(\omega)$ is obtained by demanding that the commutation relations (\ref{Ccom}) hold for the operator (\ref{Cfq}) and its hermitian conjugate. The first of (\ref{Ccom}), after elimination of $f_{\Pi_q}(\omega)$ and $f_{\Pi_X}(\omega',\omega)$ using (\ref{fPifq}) and (\ref{fPifX}), yields
\begin{equation}   \label{f*f}
\fl
(\omega+\omega')f^*_q(\omega)f_q(\omega')+(\omega+\omega')\int_{0}^{\infty}\rmd \omega'' \,f^*_X(\omega'',\omega)f_X(\omega'',\omega')=\hbar\delta(\omega-\omega').
\end{equation}
The key part of the diagonalization is the insertion of (\ref{fXsol}) into (\ref{f*f}). A product of terms containing the infinitesimal number $0^+$ is encountered, whose value depends quadratically on $0^+$. This means that the form (\ref{fXsol}) of $f_X(\omega',\omega)$ must be used in (\ref{f*f}) rather than the form (\ref{fXsol0}). Any diagonalization involving the continuum reservoir will encounter such products, and their treatment is described in detail in~\cite{phi10}, where the Hamiltonian of macroscopic QED is diagonalized. The result of substituting (\ref{fXsol}) into (\ref{f*f}) and applying (\ref{fqeffeq}) is
\begin{equation} \label{hXeq}
2\omega h^*_X(\omega)h_X(\omega)=\hbar,
\end{equation}
which has a simple solution
\begin{equation}  \label{hX}
h_X(\omega)=\left(\frac{\hbar}{2\omega}\right)^{1/2}.
\end{equation}
Other solutions of (\ref{hXeq}) differ from (\ref{hX}) by a phase factor. This freedom in the choice of the diagonalizing transformation is clear from the outset, as the diagonalizing operators (\ref{Cdef}) are only defined by (\ref{HC}) up to a phase factor. A similar analysis shows that the second commutation relation in (\ref{Ccom}) is identically satisfied by (\ref{Cfq}) because of (\ref{fXsol}) and (\ref{fqeffeq}).

We must also check that the commutation relations (\ref{cancom1}) and (\ref{cancom2}) of the original canonical operators are satisfied by the representations (\ref{qfC1})--(\ref{qfC3}). This is a consistency check on the $f$-coefficients, which have already been largely determined by (\ref{hXeq}). It is here that we find a restriction on the coupling function $\alpha(\omega)$. The full set of coupling functions that satisfies the restriction is determined by rather complicated integral relations, but for a subclass of coupling functions that includes (\ref{alpar}) the restriction is much simpler. Consider the set of functions $\alpha(\omega)$ for which
\begin{equation}   \label{alclass}
\fl
\alpha^2(\omega) \ \mathrm{is\ an\ even\ function\ of}\ \omega \qquad \mathrm{and} \qquad \frac{\alpha^2(\omega)}{\omega}>0 \ \mathrm{except\  possibly\ at}\ \omega=0.
\end{equation}
For coupling functions satisfying (\ref{alclass}) the diagonalization can be achieved if and only if
\begin{equation}
\eqalign{
\fl
\mathrm{either}  \quad \omega_0^2>\int_0^\infty\rmd \xi  \frac{\alpha^2(\xi)}{\xi^2},   \cr
\fl
 \mathrm{or}  \quad  \omega_0^2=\int_0^\infty\rmd \xi  \frac{\alpha^2(\xi)}{\xi^2} \quad \mathrm{and} \quad  \omega_0^2+\kappa^2-\int_{0}^{\infty}\rmd \xi\frac{\alpha^2(\xi)}{\xi^2+\kappa^2}\sim \kappa^n, \ n\leq 1,\quad  \mathrm{as}  \   \kappa\to 0.   } \label{alcondiag}
\end{equation}
The proof of this is given in~\ref{ap:diag}, which furthermore shows that the diagonalizing transformation requires the choice
\begin{equation} \label{hq=0}
h_q(\omega)=0
\end{equation}
as the solution of (\ref{hqeq}). It is also shown in~\ref{ap:diag} that for the class of coupling functions (\ref{alclass}), there is only one possible real root of the function in brackets in (\ref{hqeq}), at $\omega=0$. Moreover this $\omega=0$ root occurs if and only if $\omega_0^2=\int_0^\infty\rmd \xi  \alpha^2(\xi)/\xi^2$, which we recognize as the condition (\ref{alcon}) for the existence of a zero mode.  Thus, the second condition in (\ref{alcondiag}) is a restriction on coupling functions that have a zero mode.

Assuming the conditions (\ref{alclass}) and (\ref{alcondiag}) are met, a set of $f$-coefficients for the diagonalizing transformation is specified by (\ref{fXsol}), (\ref{fqsol}), (\ref{hX}) and (\ref{hq=0}). The resulting relations (\ref{qfC1})--(\ref{qfC3}) for the canonical operators in terms of the eigenmode creation and annihilation operators are most simply written in the frequency domain. With the Fourier definition
\begin{equation} \label{qopfreq}
\hat{q}(t)=\frac{1}{2\pi}\int_0^\infty\rmd\omega\left[\hat{q}(\omega)\exp(-\rmi\omega t)+\mathrm{h.c.}\right], 
\end{equation}
for $\hat{q}(\omega)$, etc., the frequency-domain canonical operators are
\begin{eqnarray}
\hat{q}(\omega)=2\pi\left(\frac{\hbar}{2\omega}\right)^{1/2}\alpha(\omega)G(\omega)\hat{C}(\omega) =\frac{\rmi}{\omega}\hat{\Pi}_q(\omega),  \label{qC} \\
\fl
\hat{X}_\omega(\omega')=2\pi\left(\frac{\hbar}{2\omega}\right)^{1/2}\delta(\omega-\omega')\hat{C}(\omega') +\frac{\alpha(\omega)}{2\omega}\left(\frac{1}{\omega-\omega'-\rmi 0^+}+\frac{1}{\omega+\omega'}\right)\hat{q}(\omega')   \label{XC} \\
=\frac{\rmi}{\omega'}\hat{\Pi}_{X_\omega}(\omega'),  \label{PiXC}
\end{eqnarray}
where the Green function $G(\omega)$ is defined in (\ref{fqsol}). The Green function can be written
\begin{equation}  \label{Gchi}
G(\omega)=\frac{-1}{\omega^2-\omega_0^2\left[1-\chi(\omega)\right]},
\end{equation}
in terms of a dimensionless quantity $\chi(\omega)$ that gives the dependence on the coupling function $\alpha(\omega)$:
\begin{equation}  \label{chi}
\omega_0^2\,\chi(\omega)=\mathrm{P}\int_{0}^{\infty}\rmd \xi\frac{\alpha^2(\xi)}{\xi^2-{\omega}^2}+\frac{\rmi\pi \alpha^2(\omega)}{2\omega}.
\end{equation}
As noted in~\ref{ap:diag}, if $\alpha^2(\omega)$ is an even function of $\omega$ then $\chi(\omega)$ exhibits a Kramers-Kronig relation between its real and imaginary parts and is therefore analytic in the upper-half complex $\omega$-plane. We can view $\chi(\omega)$ as an effective ``susceptibility" for the damped $q$-oscillator; remarkably, for couplings such that $\alpha^2(\omega)$ is even, a single damped oscillator exhibits dispersion and dissipation connected by Kramers-Kronig relations. There are in fact many similarities between the quantum damped harmonic oscillator  and macroscopic QED~\cite{phi10}, where the incorporation of Kramers-Kronig relations for electromagnetic susceptibilities is an essential part of the theory.

In sections~\ref{sec:solnI} and~\ref{sec:solnII} we solved the classical dynamics for the coupling function (\ref{alpar}). We note that this coupling function is of the class (\ref{alclass}). In addition, the function (\ref{alpar}) has the properties
\begin{eqnarray}
\int_0^\infty\rmd \xi  \frac{\alpha^2(\xi)}{\xi^2} = \int_0^\infty\rmd \xi  \frac{2\gamma\omega_0^2}{\pi(\xi^2+\gamma^2)}= \omega_0^2,   \\
\omega_0^2+\kappa^2-\int_{0}^{\infty}\rmd \xi\frac{\alpha^2(\xi)}{\xi^2+\kappa^2}=\omega_0^2+\kappa^2-\frac{\gamma\omega_0^2}{\kappa+\gamma}=\frac{\omega_0^2}{\gamma}\kappa+O(\kappa^2).
\end{eqnarray}
The second of the restrictions (\ref{alcondiag}) is thus satisfied, and therefore the Hamiltonian is diagonalizable in this case. The effective susceptibility for the coupling (\ref{alpar}) is, from (\ref{chi}),
\begin{equation} \label{chipar}
\chi(\omega)=\frac{\gamma}{\gamma-\rmi\omega}.
\end{equation}
The Green function (\ref{Gchi}) is then
\begin{equation} \label{GR2}
G_R(\omega)=-\frac{\omega+\rmi\gamma}{(\omega+\rmi0^+)(\omega^2+\rmi\gamma\omega-\omega_0^2)},
\end{equation}
where we have moved a pole at $\omega=0$ into the lower-half complex plane; this gives the retarded Green function $G_R(\omega)$, analytic in the upper-half plane. The retarded Green function (\ref{GR2}) is identical to (\ref{GRfreq}), which featured in the classical solution (as discussed above, the $f$-coefficients are a particular solution of the classical dynamical equations). The choice of the retarded Green function is not required for the diagonalization; other choices will give $f$-coefficients that differ by a zero-mode solution, described in section~\ref{sec:solnII}. 

\section{Coherent states}   \label{sec:coh}
Having diagonalizaed the quantum (and classical) Hamiltonian for coupling functions that include (\ref{alpar}), we can now construct the quantum state of the system that is closest to the interesting classical solution derived in section~\ref{sec:solnI} and plotted in Fig.~\ref{fig:parsol}. We expect the quantum states that are closest to classical solutions to be coherent states, and one benefit of the diagonalization results (\ref{qC}) and (\ref{XC}) is that they allow us to construct coherent states of the eigenmodes of the system. Classically, what we are about to do is to relate the particular solution in section~\ref{sec:solnI} to the classical normal modes of the system.

The eigenmode annihilation operator $\hat{C}(\omega)$ defines continuous-mode coherent states~\cite{mandel} by
\begin{equation}  \label{cohdef}
\hat{C}(\omega)|C(\omega)\rangle=C(\omega)|C(\omega)\rangle,
\end{equation}
where $C(\omega)$ is an arbitrary complex function. The classical solution for coupling (\ref{alpar}) gives the relations (\ref{qfreqretpar}), (\ref{qfreqsolret}) and (\ref{GRfreq}) for $q(\omega)$, while the quantum solution for $\hat{q}(\omega)$ with coupling (\ref{alpar}) is given by (\ref{qC}) with Green function (\ref{GR2}). Comparing these classical and quantum results we see that every classical solution given by the real functions $A_R(\omega)$ and $B_R(\omega)$ has a corresponding quantum coherent state with complex amplitude
\begin{equation}  \label{Ccoh}
C(\omega)=-\left(\frac{\omega}{2\hbar}\right)^{1/2}\left[A_R(\omega)+\rmi B_R(\omega)\right].
\end{equation}
The expectation values of $\hat{q}(t)$ and $\hat{X}_\omega(t)$ in the coherent states are equal to the corresponding classical solutions $q(t)$ and $X_\omega(t)$. It follows from the results at the end of section~\ref{sec:solnII} that the coherent state corresponding to the classical solution (\ref{qtparhom1})--(\ref{qtparhom3}) and (\ref{Xtparhom1})--(\ref{Xtparhom3}) has amplitude (\ref{Ccoh}) with $A_R(\omega)$ and $B_R(\omega)$ given by (\ref{ARpar}) and (\ref{BRpar}).

The expectation value of the normal-mode displacement operator $\hat{\Phi}_\omega(t)$ in a coherent state is, from (\ref{Cdef}), (\ref{cohdef}) and (\ref{Ccoh}),
\begin{eqnarray}
\langle C(\omega)|\hat{\Phi}_\omega(t) |C(\omega)\rangle&=\left(\frac{\hbar}{2\omega}\right)^{1/2}\left[C(\omega)\exp(-\rmi\omega t)+\mathrm{c.c.}\right]  \\
&=-A_R(\omega)\cos(\omega t)-B_R(\omega)\sin(\omega t).
\end{eqnarray}
This is just the general solution for the classical normal modes $\Phi_\omega(t)$ of the system. There are many other results that can be derived for the coherent states, but in this paper the quantum state we will consider in detail is the thermal state.

\section{Thermal equilibrium}  \label{sec:thermal}
Another benefit of diagonalizing the Hamiltonian is that it allows a straightforward calculation of the thermodynamic quantities for the $q$-oscillator. In this section we treat the case of thermal equilibrium for all couplings for which the diagonalization in section~\ref{sec:quantization} is valid;  the results for the particular coupling (\ref{alpar}), corresponding to damping proportional to velocity, will also be described. Given the close relationship between the damped harmonic oscillator and macroscopic QED, noted in the last section, it is not surprising that the following treatment of the thermal state of the damped oscillator has similarities to the Casimir effect, which is simply macroscopic QED in thermal equilibrium~\cite{phi11}.

In thermal equilibrium the excitation level of each eigenmode oscillator of the system will be given by the Planck distribution, so the thermal mixed state is defined by
\begin{eqnarray}
\langle\hat{C}^\dagger(\omega)\hat{C}(\omega')\rangle=\mathcal{N}(\omega)\delta(\omega-\omega')=\langle\hat{C}(\omega)\hat{C}^\dagger(\omega')\rangle-\delta(\omega-\omega'),   \label{CCdthermal} \\
\mathcal{N}(\omega)=\left[\exp\left(\frac{\hbar\omega}{k_BT}\right)-1\right],\label{planck}   \\
\langle\hat{C}(\omega)\hat{C}(\omega')\rangle=0.  \label{CCthermal}
\end{eqnarray}
The continuum of eigenmodes necessitates the use of delta functions in (\ref{CCdthermal}). These delta functions lead to some awkwardness in the thermal calculations, but as in~\cite{phi11} such difficulties can be negotiated by regarding such delta functions as limits to be strictly imposed only at the end of calculations. Using (\ref{CCdthermal})--(\ref{CCthermal}) it is straightforward to compute the thermal position and momentum correlation functions for the $q$-oscillator, as well as its thermal (including zero-point) energy and other thermodynamic quantities. Our procedure will be to derive the results for a general coupling function for which the diagonalization in section~\ref{sec:quantization} is valid, and then to examine those results for the particular coupling (\ref{alpar}). 

\subsection{Thermal correlation functions for the $q$-oscillator}  \label{sec:qthermal}
It follows from (\ref{qopfreq}), (\ref{qC}) and (\ref{CCthermal}) that the thermal correlation function of $\hat{q}(t)$ can be written
\begin{eqnarray}
\fl
\langle \hat{q}(t)\hat{q}(t')\rangle=\frac{1}{4\pi^2}\int_0^\infty\rmd\omega\int_0^\infty\rmd\omega' & \left\{   \exp\left[-\rmi(\omega t-\omega't')\right]\langle \hat{q}(\omega)\hat{q}^\dagger(\omega')\rangle  \right.  \nonumber \\
&  \left. +  \exp\left[\rmi(\omega t-\omega't')\right]\langle \hat{q}^\dagger(\omega)\hat{q}(\omega')\rangle \right\},        \label{qqtime0}
\end{eqnarray}
with a similar result for the other canonical operators of the system. The frequency-domain correlation functions $\langle \hat{q}^\dagger(\omega)\hat{q}(\omega')\rangle$ and $\langle \hat{q}(\omega)\hat{q}^\dagger(\omega')\rangle$ are, from (\ref{qC}) and (\ref{CCdthermal}),
\begin{equation}   \label{qdq0}
\eqalign{ \langle \hat{q}^\dagger(\omega)\hat{q}(\omega')\rangle& =\frac{2\pi^2\hbar\alpha^2(\omega)}{\omega}G^*(\omega)G(\omega')\mathcal{N}(\omega)\delta(\omega-\omega') \cr
& =\frac{\mathcal{N}(\omega)}{\mathcal{N}(\omega)+1}\langle \hat{q}(\omega)\hat{q}^\dagger(\omega')\rangle.       }
\end{equation}
Multiplying the Green function expression (\ref{Gchi}) by $G^*(\omega)$, we obtain
\begin{equation}  \label{G*G}
\left\{{\omega}^2-\omega_0^2\left[1-\chi(\omega)\right]\right\}G^*(\omega)G(\omega)=-G^*(\omega),
\end{equation}
and the imaginary part of this equation is found using (\ref{chi}):
\begin{equation}  \label{ImG}
\frac{\pi\alpha^2(\omega)}{2\omega}G^*(\omega)G(\omega)=\mathrm{Im}G(\omega).
\end{equation}
Inserting (\ref{ImG}) in (\ref{qdq0}) we find
\begin{equation}   \label{ddq}
\eqalign{ \langle \hat{q}^\dagger(\omega)\hat{q}(\omega')\rangle & =4\pi\hbar\delta(\omega-\omega')\mathcal{N}(\omega)\mathrm{Im}G(\omega) \cr
& =\frac{\mathcal{N}(\omega)}{\mathcal{N}(\omega)+1}\langle \hat{q}(\omega)\hat{q}^\dagger(\omega')\rangle,       }
\end{equation}
which shows that the real part of the temporal correlation function (\ref{qqtime0}) is
\begin{equation}  \label{qqtime}
\fl
\frac{1}{2}\left\langle \hat{q}(t)\hat{q}(t') + \hat{q}(t')\hat{q}(t)\right\rangle=\frac{\hbar}{\pi}\int_0^\infty \rmd\omega \,\cos\left[\omega(t-t')\right]\coth\left(\frac{\hbar\omega}{2k_BT}\right) \mathrm{Im}G(\omega),
\end{equation}
where $2\mathcal{N}(\omega)+1$ has been rewritten as a hyperbolic tangent. The momentum operator for the $q$-oscillator is given by (\ref{qC}) and is just the time derivative of $\hat{q}(t)$, so we obtain from (\ref{qqtime}) the momentum correlation function
\begin{equation}  \label{PqPqtime}
\fl
\frac{1}{2}\left\langle \hat{\Pi}_q(t)\hat{\Pi}_q(t') + \hat{\Pi}_q(t')\hat{\Pi}_q(t)\right\rangle=\frac{\hbar}{\pi}\int_0^\infty  \!\!  \rmd  \omega \,\omega^2\cos\left[\omega(t-t')\right]\coth\left(\frac{\hbar\omega}{2k_BT}\right) \mathrm{Im}G(\omega).
\end{equation}

We now consider the general results (\ref{qqtime}) and (\ref{PqPqtime}) in the case of the coupling function (\ref{alpar}), corresponding to damping of the $q$-oscillator proportional to velocity. We first note a slight modification of the result (\ref{ImG}) in the case of (\ref{alpar}). Equation~(\ref{G*G}) does not include a possible pole prescription at $\omega=0$ for a zero mode, such as occurs in (\ref{GR2}). Evaluating the left-hand side of (\ref{ImG}) for the retarded Green function (\ref{GR2}), and replacing $0^+$ by $\eta$, we find
\begin{equation}  \label{ImGpar}
\frac{\pi\alpha^2(\omega)}{2\omega}G_R^*(\omega)G_R(\omega)=\frac{\gamma\omega\omega_0^2}{(\omega^2+\eta^2)[(\omega^2-\omega_0^2)^2+\gamma^2\omega^2]}.
\end{equation}
As $\eta\to 0$, the right-hand side of (\ref{ImGpar}) has a pole at $\omega=0$ and $\eta$ prescribes the treatment of this pole by a principal value; apart from the pole prescriptions, (\ref{ImGpar}) is the result (\ref{ImG}). With use of (\ref{ImGpar}) the correlation function (\ref{qqtime}) gives the following expectation value $\langle\hat{q}^2(t)\rangle$ for coupling (\ref{alpar}):
\begin{equation}  \label{qqpar}
\left\langle \hat{q}^2(t)\right\rangle=\frac{\hbar}{2\pi}\int_{-\infty}^\infty \rmd\omega \,\frac{\gamma\omega\omega_0^2}{(\omega^2+\eta^2)[(\omega^2-\omega_0^2)^2+\gamma^2\omega^2]} \coth\left(\frac{\hbar\omega}{2k_BT}\right).
\end{equation}
In (\ref{qqpar}) the integration has been extended over negative $\omega$ (the integrand is an even function of $\omega$); this allows evaluation of the integral by closing the integration contour in the upper (or lower) half complex plane. It is clear that the integral in (\ref{qqpar}) does not converge when $\eta=0$, and evaluation by contour integration for $\eta>0$ shows that this divergence is given by a term $k_BT\gamma/(\omega_0^2\eta)$. The form of the divergent term shows that (\ref{qqpar}) is finite at $T=0$, so that the ground state value of $\langle\hat{q}^2(t)\rangle$ is well defined. The divergence at $T>0$ can be attributed to the zero mode, as it is associated with the $\omega=0$ pole in the Green function. As the zero mode has zero energy, it is perhaps not surprising that the zero-mode squared displacement diverges at non-zero temperature. Although it is tempting to drop the divergent zero-mode term $k_BT\gamma/(\omega_0^2\eta)$ from (\ref{qqpar}) when $T>0$, this would be a violation of quantum mechanics because the canonical commutation relations of the system only hold for the full displacement operator $\hat{q}(t)$. There is thus an inherent pathology in the thermal state for the coupling (\ref{alpar}), although the ground state is well-defined as are other quantum states such as the infinite class of coherent states described in section~\ref{sec:coh}. Note that the divergent term $k_BT\gamma/(\omega_0^2\eta)$ also vanishes if the limit $\gamma\to 0$ in the integration result is taken before the limit $\eta\to0$; in this case the result for $\langle\hat{q}^2(t)\rangle$ reduces to the free-oscillator value $\frac{\hbar}{2\omega_0}\coth\left(\frac{\hbar\omega_0}{2k_BT}\right)$. As discussed in section~\ref{sec:quantization}, coupling functions in the class (\ref{alclass}) give rise to a zero-mode only if they obey the special condition (\ref{alcon}); for general coupling functions in the class (\ref{alclass}) the correlation function (\ref{qqtime}) will be well behaved. The momentum-squared expectation value $\langle\hat{\Pi}_q^2(t)\rangle$ for the coupling (\ref{alpar}) is found from (\ref{PqPqtime}) and (\ref{ImGpar}) to be
\begin{equation}  \label{PqPqpar}
\left\langle\hat{\Pi}_q^2(t)\right\rangle=\frac{\hbar}{2\pi}\int_{-\infty}^\infty \rmd\omega \,\frac{\gamma\omega\omega_0^2}{(\omega^2-\omega_0^2)^2+\gamma^2\omega^2} \coth\left(\frac{\hbar\omega}{2k_BT}\right),
\end{equation}
where there is no longer any contribution of the $\omega=0$ pole in the Green function. The integral in (\ref{PqPqpar}) is well defined and can be evaluated by contour methods but, given the pathology in the corresponding position quantity $\langle\hat{q}^2(t)\rangle$ at $T>0$, the rather complicated details are omitted here. We note however that when the the limit $\gamma\to0$ is taken in the final result, (\ref{PqPqpar}) gives the free-oscillator value $\frac{1}{2}\hbar\omega_0\coth\left(\frac{\hbar\omega_0}{2k_BT}\right)$.

\subsection{Reservoir contributions to thermal correlation functions}
The thermal correlation functions of the reservoir are of interest because of their contribution to the thermal energy of the $q$-oscillator. As discussed in section~\ref{sec:qenergy}, for this purpose we must calculate correlation functions containing the reservoir using (\ref{XC}) and (\ref{PiXC}), but only retain terms that contain $\hat{q}(\omega)$. This is the prescription used in macroscopic QED to obtain the Casimir stress-energy of the electromagnetic field in a medium, where the electric and magnetic fields play the role of the $q$-oscillator~\cite{phi11}. In fact the results in this subsection are largely a simpler version of the reservoir correlation functions in~\cite{phi11}.

The expectation value of the Hamiltonian (\ref{H}) in thermal equilibrium contains the term $\int_0^\infty\rmd\omega\langle(\partial_t\hat{X}_\omega)^2+\omega^2\hat{X}_\omega^2\rangle/2$. We compute this expectation value using (\ref{XC}), retaining only terms in which $\hat{q}(\omega)$ occurs. Essentially the same calculation is described in detail in~\cite{phi11} for macroscopic QED. We therefore refer the reader to~\cite{phi11} for the various steps in the derivation and here state the result:
\begin{equation}   \label{HXX}
\fl
\frac{1}{2}\int_0^\infty\rmd\omega\left\langle(\partial_t\hat{X}_\omega)^2+\omega^2\hat{X}_\omega^2\right\rangle=\frac{\hbar\omega_0^2}{2\pi}\mathrm{Im}\int_0^\infty\rmd\omega \coth\left(\frac{\hbar\omega}{2k_BT}\right)\frac{\rmd\left[\omega\chi(\omega)\right]}{\rmd\omega}G(\omega).
\end{equation}
The thermal expectation value of the Hamiltonian (\ref{H}) also contains $\int_0^\infty\rmd\omega\,\alpha(\omega)\langle \hat{q}X_\omega\rangle$. This expectation value is also found using (\ref{XC}), but here $\hat{q}(\omega)$ occurs in all contributions to the expectation value, so no terms are dropped. There is again an essentially identical calculation in~\cite{phi11} for macroscopic QED, and the same derivation gives the result
\begin{equation}  \label{HqX}
\fl
\int_0^\infty\rmd\omega\,\alpha(\omega)\left\langle \hat{q}(t)\hat{X}_\omega(t)\right\rangle=\frac{\hbar\omega_0^2}{\pi}\mathrm{Im}\int_0^\infty\rmd\omega \coth\left(\frac{\hbar\omega}{2k_BT}\right)\chi(\omega)G(\omega).
\end{equation}
This quantity is real so the term containing $(\hat{q}\hat{X}_\omega+\hat{X}_\omega\hat{q})/2$ in the Hamiltonian (\ref{H}) gives the thermal expectation value on the right-hand side of (\ref{HqX}). The thermal (including zero-point) energy of the $q$-oscillator can now be obtained.

\subsection{Thermal energy of the $q$-oscillator} \label{sec:qenergy}
The energy of the $q$-oscillator in any state is given by the expectation value of the Hamiltonian (\ref{H}) with the reservoir $X_\omega$ traced out. Due to the coupling of $q$ to $X_\omega$, this tracing operation is in general not a trivial matter. The same issue arises in macroscopic QED, where the value of the electromagnetic energy-momentum tensor in a dispersive, dissipative medium requires the tracing out of the reservoir. In the case of thermal equilibrium the electromagnetic stress-energy was found in~\cite{phi11} to be given by a straightforward recipe: after the reservoir operators are expressed in terms of electromagnetic-field operators, all contributions containing electromagnetic-field operators are to be retained, the other contributions are to be dropped. We follow the same recipe here to trace out the reservoir from the total energy of the system in thermal equilibrium and thereby obtain the thermal energy of the $q$-oscillator. The reservoir operators are expressed in terms of the $q$-operator using  (\ref{XC}) and (\ref{PiXC}), and the thermal expectation value of the Hamiltonian (\ref{H}) is evaluated with only contributions that contain $\hat{q}(\omega)$ included. The result is the thermal energy of the $q$-oscillator, which we denote by $\langle\hat{H}\rangle_q$. The required expectation values involving the reservoir are given by (\ref{HXX}) and (\ref{HqX}), and the thermal expectation values of the first two terms in the Hamiltonian (\ref{H}) are found from (\ref{qqtime}) and (\ref{PqPqtime}). This gives the following expression for $\langle\hat{H}\rangle_q$:
\begin{equation}   \label{Hq}
\fl
\langle\hat{H}\rangle_q=\frac{\hbar}{4\pi}\mathrm{Im}\int_{-\infty}^\infty\rmd\omega \coth\left(\frac{\hbar\omega}{2k_BT}\right)\left\{\omega_0^2\left[\omega\frac{\rmd\chi(\omega)}{\rmd\omega}-\chi(\omega)+1\right]+\omega^2\right\}G(\omega).
\end{equation}
Due to the parity properties of the susceptibility and the Green function, the real part of the integrand in (\ref{Hq}) does not contribute and the integral is imaginary. The $q$-oscillator thermal energy (\ref{Hq}) is the analogue for the damped oscillator of the Casimir energy density of electromagnetic fields in a medium~\cite{phi11}. A dispersive contribution to the energy from the derivative of the susceptibility is familiar in the electromagnetic case~\cite{phi11}, and we find a similar contribution from the damped-oscillator ``susceptibility" in (\ref{Hq}). As is familiar in Casimir theory and elsewhere, the integral in (\ref{Hq}) can be converted to a sum over imaginary Matsubara frequencies.

In the case of the coupling (\ref{alpar}) with susceptibility (\ref{chipar}), the real part of the integrand in (\ref{Hq}) does not give a convergent integral so that it is not correct to take the Im outside the integration as has been done in the derivation of (\ref{Hq}). For the coupling (\ref{alpar}) we must keep the Im inside the integration and take the imaginary part of the integrand; the resulting expression for $\langle\hat{H}\rangle_q$ is
\begin{equation}  \label{Hqpar}
\langle\hat{H}\rangle_q=\frac{\hbar}{4\pi}\int_{-\infty}^\infty \rmd\omega \,\frac{\gamma\omega\omega_0^2(\gamma^2+3\omega^2-\omega_0^2)}{(\omega^2+\gamma^2)\left[(\omega^2-\omega_0^2)^2+\gamma^2\omega^2\right]} \coth\left(\frac{\hbar\omega}{2k_BT}\right).
\end{equation}
There is no zero-mode contribution in (\ref{Hqpar}), as is expected from the fact that the zero mode has zero energy. The thermal energy (\ref{Hqpar}) is finite and can be evaluated analytically by closing the integration contour in the upper (or lower) half complex plane. When the zero-damping limit $\gamma\to0$ is taken, however, the result for the thermal energy is found to reduce to
\begin{equation}   \label{Hqfree}
\frac{1}{2}\hbar\omega_0\coth\left(\frac{\hbar\omega_0}{2k_BT}\right)-\frac{1}{2}k_BT,
\end{equation}
i.e. in addition to the free-oscillator thermal energy there is an additional term $-k_BT/2$. This additional term can be traced back to the dispersive contribution $\rmd\chi(\omega)/\rmd\omega$ in (\ref{Hq}); although $\chi(\omega)$ vanishes as $\gamma\to0$, the contribution from this dispersive term to the energy  becomes $-k_BT/2$ as $\gamma\to0$. The non-zero contribution of the $\rmd\chi(\omega)/\rmd\omega$-term when $\gamma\to0$ relies on the assumption that the susceptibility (\ref{chipar}) is valid up to infinite frequencies. Although it is mathematically convenient to employ functions like (\ref{chipar}) over the entire frequency range, this is an unphysical assumption and therefore may produce some unphysical results. As will be discussed in section~\ref{sec:concl}, application of the general results presented here to real physical systems would probably require a measurement of the susceptibility characterizing the particular damped oscillator; although these measured susceptibilities could be fitted to mathematical functions, non-zero values of these functions at very large frequencies beyond the measured range would be physically meaningless. We note that the difficulty with the zero-coupling limit of the thermal energy for damping (\ref{alpar}) does not occur for the ground state $T=0$, as we see from (\ref{Hqfree}) that the correct free-oscillator limit $\hbar\omega_0/2$ is obtained. These thermal-energy results support the comment in section~\ref{sec:qthermal} that the coupling (\ref{alpar}) gives an inherent pathology in the non-zero temperature state.

\section{Conclusions}  \label{sec:concl}
The use of a discrete set of oscillators as a reservoir is almost universal in studying dissipation in quantum systems using exact quantization rules. Yet the power of a continuum reservoir in the study of dissipation was demonstrated twenty years ago by Huttner and Barnett~\cite{hut92}, and it has recently been shown that a continuum reservoir allows an exact canonical quantization of the macroscopic Maxwell equations in arbitrary media obeying Kramers-Kronig relations~\cite{phi10,phi11,hor11,hor12}. The important results that have been obtained with a continuum reservoir would not have been achieved if a limiting procedure had been applied to results featuring a discrete reservoir, whereas the discrete reservoir almost never captures the desired physics without a subsequent, and delicate, limit to the continuum case~\cite{tat87}. In view of this, it is highly advisable to employ a continuum reservoir from the outset. We have considered a single damped quantum harmonic oscillator and shown that a continuum reservoir allows the exact treatment of damping proportional to velocity. In addition we have found that for general damping the quantum oscillator is remarkably similar to light in macroscopic media. For functions $\alpha(\omega)$ coupling the oscillator to the reservoir, with $\alpha^2(\omega)$ an even function of $\omega$, the dynamics of the damped oscillator is governed by an effective susceptibility obeying Kramers-Kronig relations.

The experimental investigation of macroscopic oscillators that exhibit quantum behaviour has made remarkable progress in recent years; oscillators have been cooled to their quantum ground states and even placed in a superposition of energy states~\cite{oco10,teu11,cha11,asp10,poo12}. A theoretical description of nano-oscillators and opto-mechanical systems can be approached in various ways, depending among other things on how much detail of the microscopic structure of the oscillator is included. The approach to the damped oscillator taken in this paper naturally leads to a description of the oscillator by means of an effective susceptibility. As in macroscopic electromagnetism, the strength of this approach is that the complicated microscopic substructure is not included, rather the susceptibility is a quantity to be measured for real physical systems. We have found that diagonalization of the Hamiltonian requires restrictions on the coupling function that determines the susceptibility, but this may not be significant for real macroscopic oscillators. The condition (\ref{alcondiag}) constrains the relationship between the coupling function $\alpha(\omega)$ and the free-oscillation frequency $\omega_0$. But most mechanical oscillators do not exhibit a ``free" oscillation state; the oscillation degree of freedom is produced by the same material geometry that causes damping. This means that the parameter $\omega_0$ cannot be separated from the susceptibility, and they must together be fitted to a given system through experimental measurement. A choice of $\omega_0$ for which $\omega_0^2>\int_0^\infty\rmd\omega\,\alpha^2(\omega)/\omega^2$ may therefore not be restrictive. This also suggests that the there is freedom in the form of the coupling term between the $q$-oscillator and the reservoir. We chose a coupling of the form $qX_\omega$, whereas Huttner and Barnett~\cite{hut92} use $q\dot{X}_\omega$, and the latter does not give the restriction (\ref{alcondiag}) on the coupling function. But the Huttner-Barnett coupling leads to a renormalization of the free-oscillation frequency $\omega_0$~\cite{hut92}, which is why the restriction (\ref{alcondiag}) is avoided. If $\omega_0$ has no direct physical meaning one can effectively make the same redefinition with the coupling term chosen here, through the joint characterization of $\omega_0$ and the susceptibility $\chi(\omega)$.

The considerations in this paper are of course only a first step towards describing physical quantum oscillators, and other ingredients would have to be included such as coupling to light in the case of opto-mechanical systems. There is also the crucial question of how the effective susceptibility is to be measured for real oscillators, or how physically realistic susceptibilities could be deduced from simple microscopic considerations, as is done in electromagnetism. In regard to this last point, damping proportional to velocity would not be expected to be physically relevant. While the dynamics (\ref{qdampf}) may be a reasonable description, for small $\dot{q}$, of a free oscillator in a fluid, nano-oscillators with mechanical damping determined by the particular material geometry would not experience a simple damping proportional to velocity. Real macroscopic oscillators may however be describable by an effective susceptibility and by results such as the thermal energy (\ref{Hq}). 

\ack
I am indebted to Simon Horsley for many helpful disscussions. I also thank Gabriel Barton for useful literature and information on several aspects of this work. This research is supported by the Royal Society of Edinburgh and the Scottish Government.

\appendix
\section{Green function} \label{ap:green}
Here we present a Green function $G(t,t_0)$ that satisfies (\ref{Gpareq}). We obtain the solution by solving (\ref{Ginteq}) for $G(\omega,t_0)$ with $\epsilon>0$. The Green function corresponds to an infinitesimal value of $\epsilon$, but (\ref{Ginteq}) is first solved for a non-infinitesmal number $\epsilon>0$ for which the standard solution method~\cite{pipkin} can be applied without difficulty; after the solution is obtained the limit of infinitesimal $\epsilon$ is taken and then $G(\omega,t_0)$ is (inverse) Fourier transformed to find $G(t,t_0)$. The solution of (\ref{Ginteq}) with $\epsilon>0$ differs depending on whether $t_0$ is positive or negative, as this affects the analytic properties of the inhomogeneous term as a function of $\omega$. The general solution of (\ref{Ginteq}) will also contain the solution (\ref{qfreqparhom}) of the homogeneous integral equation (\ref{qinteqparhom}), which we do not require here and therefore drop. We find the solution $G(\omega,t_0)$ by the standard technique~\cite{pipkin} and then take $\epsilon\to 0^+$, whereupon the infinitesimal number $0^+$ serves only to regulate a pole at $\omega=0$:
\begin{eqnarray}
\fl
G(\omega,t_0)=-\frac{(\omega+\rmi\gamma)\exp(\rmi\omega t_0)}{(\omega-\rmi 0^+)(\omega^2-\omega_0^2+\rmi\gamma\omega)}-\frac{2\rmi\gamma\omega_0^2}{(\omega-\rmi 0^+)[(\omega^2-\omega_0^2)^2+\gamma^2\omega^2]}   \nonumber \\[3pt]
\fl
\qquad\quad+\frac{\exp\left(-\frac{\gamma t_0}{2}\right)\left[(\omega_0^2-\gamma^2)\omega_1\cos\left(\frac{\omega_1t_0}{2}\right)-\gamma(\gamma^2-3\omega_0^2)\sin\left(\frac{\omega_1t_0}{2}\right)\right]}{\omega_1[(\omega^2-\omega_0^2)^2+\gamma^2\omega^2]}, \qquad t_0\geq 0,  \label{Gfreq1} \\[3pt]
\fl
G(\omega,t_0)=-\frac{(\omega-\rmi\gamma)\exp(\rmi\omega t_0)}{(\omega-\rmi 0^+)(\omega^2-\omega_0^2-\rmi\gamma\omega)}   \nonumber \\[3pt]
\fl
\qquad\quad+\frac{\exp\left(\frac{\gamma t_0}{2}\right)\left[(\omega_0^2-\gamma^2)\omega_1\cos\left(\frac{\omega_1t_0}{2}\right)+\gamma(\gamma^2-3\omega_0^2)\sin\left(\frac{\omega_1t_0}{2}\right)\right]}{\omega_1[(\omega^2-\omega_0^2)^2+\gamma^2\omega^2]}, \qquad t_0\leq 0. \label{Gfreq2}
\end{eqnarray}
The solution (\ref{Gfreq1})--(\ref{Gfreq2}) is continuous at $t_0=0$, and $\omega_1$ is again given by (\ref{omega1}). Fourier transformation of  (\ref{Gfreq1})--(\ref{Gfreq2}) to the time domain gives the Green function
\begin{eqnarray}
\fl
G(t,t_0)=\left[2\theta(-t)-\theta(t_0-t)\right]\frac{\gamma}{\omega_0^2}e^{0^+t}
 \nonumber \\
\fl
\qquad   +\frac{\exp\left(-\frac{\gamma |t|}{2}\right)}{\omega_1\omega_0^2}\left[\gamma\omega_1\mathrm{sgn}(t)\cos\left(\frac{\omega_1 t}{2}\right)+(\gamma^2-2\omega_0^2)\sin\left(\frac{\omega_1 t}{2}\right) \right]  \nonumber \\[3pt]
\fl
\qquad -\theta(t-t_0)\frac{\exp\left[-\frac{\gamma}{2}(t-t_0)\right]}{\omega_1\omega_0^2}\left\{\gamma\omega_1\cos\left[\frac{\omega_1}{2}(t-t_0)\right]+(\gamma^2-2\omega_0^2)\sin\left[\frac{\omega_1}{2}(t-t_0)\right] \right\}   \nonumber \\[3pt]
\fl
\qquad -\frac{\exp\left[-\frac{\gamma}{2}(|t|+t_0)\right]}{2\gamma\omega_1^2\omega_0^2}  \left[\omega_1\cos\left(\frac{\omega_1 t}{2}\right)+\gamma\sin\left(\frac{\omega_1 |t|}{2}\right) \right]      \nonumber \\[3pt]
\fl
\qquad \times \left[(\gamma^2-\omega_0^2)\omega_1\cos\left(\frac{\omega_1 t_0}{2}\right)+(\gamma^2-3\omega_0^2)\gamma\sin\left(\frac{\omega_1 t_0}{2}\right)\right] , \qquad t_0\geq 0,  \label{G1} \\[3pt]
\fl
G(t,t_0)=\theta(t_0-t)\frac{\gamma}{\omega_0^2}e^{0^+t}
 \nonumber \\
 \fl
\qquad -\theta(t_0-t)\frac{\exp\left[\frac{\gamma}{2}(t-t_0)\right]}{\omega_1\omega_0^2}\left\{\gamma\omega_1\cos\left[\frac{\omega_1}{2}(t-t_0)\right]-(\gamma^2-2\omega_0^2)\sin\left[\frac{\omega_1}{2}(t-t_0)\right] \right\}   \nonumber \\[3pt]
\fl
\qquad -\frac{\exp\left[-\frac{\gamma}{2}(|t|-t_0)\right]}{2\gamma\omega_1^2\omega_0^2}  \left[\omega_1\cos\left(\frac{\omega_1 t}{2}\right)+\gamma\sin\left(\frac{\omega_1 |t|}{2}\right) \right]      \nonumber \\[3pt]
\fl
\qquad \times \left[(\gamma^2-\omega_0^2)\omega_1\cos\left(\frac{\omega_1 t_0}{2}\right)-(\gamma^2-3\omega_0^2)\gamma\sin\left(\frac{\omega_1 t_0}{2}\right)\right] , \qquad t_0\leq 0,  \label{G2} 
\end{eqnarray}
where $\theta(x)$ is the step function. The values of $G(t,t_0)$ at zeros of the step functions in (\ref{G1})--(\ref{G2}) are to be taken as limits; $G(t,t_0)$ is in fact continuous across $t=0$, $t_0=0$ and $t=t_0$. It is straightforward to verify that  (\ref{G1})--(\ref{G2}) satisfies (\ref{Gpareq}); there is a discontinuity in $\rmd G(t,t_0)/\rmd t$ at $t=t_0$ which gives a delta function in $\rmd^2 G(t,t_0)/\rmd t^2$.
The infinitesimal number $0^+$ in (\ref{G1})--(\ref{G2}) regularizes its Fourier transform in $t$, giving the $G(\omega,t_0)$ of (\ref{Gfreq1})--(\ref{Gfreq2}).

\section{Coupling functions for a diagonalizable Hamiltonian} \label{ap:diag}
Here we complete the final steps in diagonalizing the Hamiltonian. In section~\ref{sec:quantization} we constructed a transformation, given by (\ref{qfC1})--(\ref{qfC3}), (\ref{fXsol}), (\ref{fqsol}) and (\ref{hX}), that will diagonalize the Hamiltonian provided the commutation relations (\ref{cancom1}) and (\ref{cancom2}) are satisfied by (\ref{qfC1})--(\ref{qfC3}). Inserting  (\ref{qfC1})--(\ref{qfC3}) into (\ref{cancom1}) and (\ref{cancom2}), and using (\ref{Ccom}), we obtain the final non-trivial conditions on the $f$-coefficients for the diagonalizing transformation:
\begin{eqnarray}
\int_0^\infty\rmd\omega f_q(\omega) f^*_{\Pi_q}(\omega)-\int_0^\infty\rmd\omega f^*_q(\omega) f_{\Pi_q}(\omega)=\rmi\hbar,   \label{fcon1} \\
\fl
\int_0^\infty\rmd\omega'' f_X(\omega,\omega'') f^*_{\Pi_X}(\omega',\omega'')-\int_0^\infty\rmd\omega'' f^*_X(\omega,\omega'') f_{\Pi_X}(\omega',\omega'')=\rmi\hbar\delta(\omega'-\omega''),   \label{fcon2}   \\
\fl
\int_0^\infty\rmd\omega'' f_X(\omega,\omega'') f^*_{X}(\omega',\omega'')-\int_0^\infty\rmd\omega'' f^*_X(\omega,\omega'') f_{X}(\omega',\omega'')=0,  \\
\fl
\int_0^\infty\rmd\omega'' f_{\Pi_X}(\omega,\omega'') f^*_{\Pi_X}(\omega',\omega'')-\int_0^\infty\rmd\omega'' f^*_{\Pi_X}(\omega,\omega'') f_{\Pi_X}(\omega',\omega'')=0.
\end{eqnarray}
The last two of these conditions are easily seen to hold because of (\ref{fXsol}) and the first of (\ref{fPifX}). We proceed to show that the first two conditions, i.e.\ (\ref{fcon1}) and (\ref{fcon2}), hold for coupling functions satisfying (\ref{alclass}) and (\ref{alcondiag}), a class that includes the coupling function (\ref{alpar}) corresponding to damping proportional to velocity. The following analysis has an exact counterpart in the Huttner-Barnett model~\cite{hut92}, although the results are different due to a different type of coupling between the $q$-oscillator and the reservoir.

Consider first the condition (\ref{fcon1}). If we insert the first of (\ref{fPifq}) together with (\ref{fqsol}) into (\ref{fcon1}), we encounter a term quadratic in $h_q(\omega)$. As noted after (\ref{hqeq}), a non-zero $h_q(\omega)$ will contain a delta function, so to avoid a square of a delta function in (\ref{fcon1}) we are forced to choose the solution
\begin{equation}  \label{hq0}
h_q(\omega)=0
\end{equation}
of (\ref{hqeq}). The condition (\ref{fcon1}) is then, with use of (\ref{hX}),
\begin{equation}
1=\int_0^\infty\rmd\omega\,\alpha^2(\omega)\left|{\omega}^2-\omega_0^2+\mathrm{P}\int_{0}^{\infty}\rmd \xi\frac{\alpha^2(\xi)}{\xi^2-{\omega}^2}+\frac{\rmi\pi \alpha^2(\omega)}{2\omega}\right|^{-2}.      \label{alcongen}
\end{equation}
This complicated restriction on the coupling function $\alpha(\omega)$ is a requirement for the Hamiltonian to be diagonalizable. For a very broad class of coupling functions however, (\ref{alcongen}) reduces to a much simpler requirement. If $\alpha^2(\omega)$ is an even function of $\omega$ then (\ref{alcongen}) can be written
\begin{equation}
1=\frac{1}{\rmi\pi}\int_{-\infty}^\infty\rmd\omega\,\omega\left({\omega}_0^2-\omega^2-\mathrm{P}\int_{0}^{\infty}\rmd \xi\frac{\alpha^2(\xi)}{\xi^2-{\omega}^2}-\frac{\rmi\pi \alpha^2(\omega)}{2\omega}\right)^{-1}.      \label{alevencon}
\end{equation}
The assumption that $\alpha^2(\omega)$ is an even function also implies that the combination
\begin{equation}  \label{sus}
\mathrm{P}\int_{0}^{\infty}\rmd \xi\frac{\alpha^2(\xi)}{\xi^2-{\omega}^2}+\frac{\rmi\pi \alpha^2(\omega)}{2\omega},
\end{equation}
which appears in the denominator in (\ref{alevencon}), is analytic on the upper-half complex $\omega$-plane. This follows from the familiar analysis of susceptibilities that are analytic on the  upper-half frequency plane and that therefore obey Kramers-Kronig relations~\cite{LL}; the real and imaginary parts of (\ref{sus}) exhibit one of the Kramers-Kronig relations for a complex susceptibility, provided $\alpha^2(\omega)$ is an even function of $\omega$. If we further assume that $\alpha^2(\omega)/\omega>0$ except possibly at $\omega=0$, then it is shown in~\cite{LL} that the ``susceptibility" (\ref{sus}) does not take real values at any finite point in the upper-half complex $\omega$-plane except on the imaginary axis, where it varies monotonically from its value at $\rmi 0$ to its value at $\rmi\infty$. Moreover, the (real) value of (\ref{sus}) on the positive imaginary axis $\omega=\rmi\kappa$, $\kappa\geq0$, is~\cite{LL}
\begin{equation}  \label{susim}
\int_{0}^{\infty}\rmd \xi\frac{\alpha^2(\xi)}{\xi^2+\kappa^2}.
\end{equation}
We now consider the analytic properties of the integrand in (\ref{alevencon}) in the upper-half frequency plane. As (\ref{sus}) is real in the upper-half plane only on the positive imaginary axis, the denominator in (\ref{alevencon}) can vanish only at positive imaginary frequencies $\omega=\rmi\kappa$, where it has the value (using (\ref{susim}))
\begin{equation}  \label{denim}
\omega_0^2+\kappa^2-\int_{0}^{\infty}\rmd \xi\frac{\alpha^2(\xi)}{\xi^2+\kappa^2}.
\end{equation}
It is clear that (\ref{denim}) has a zero if and only if
\begin{equation}  \label{denzero}
\omega_0^2\leq \int_{0}^{\infty}\rmd \xi\frac{\alpha^2(\xi)}{\xi^2}.
\end{equation}
If the equality holds in (\ref{denzero}) then (\ref{denim}) has a zero at $\omega=\rmi\kappa=0$, but this does not give a pole in the integrand in (\ref{alevencon}) unless (\ref{denim}) goes as $\kappa^n$, $n>1$, as $\kappa\to 0$. There are therefore no poles in the integrand in (\ref{alevencon}) in the upper-half plane if and only if
\begin{equation}
\eqalign{
\mathrm{either}  \quad \omega_0^2>\int_0^\infty\rmd \xi  \frac{\alpha^2(\xi)}{\xi^2},    \cr 
 \mathrm{or} \quad  \omega_0^2=\int_0^\infty\rmd \xi  \frac{\alpha^2(\xi)}{\xi^2} \quad \mathrm{and\ \  (\ref{denim})} \sim \kappa^n, \quad n\leq 1, \quad  \mathrm{as}  \quad   \kappa\to 0. } \label{alconapp}
\end{equation}
If (\ref{alconapp}) holds then the integral in (\ref{alevencon}) is equal to minus the integral over an infinite semi-circle in the upper-half plane. Taking $\omega=Re^{\rmi\phi}$, $R\to\infty$, $0\leq\phi\leq\pi$ as the integration parameter along the infinite semi-circle, we find that the integral in (\ref{alevencon}) is $\rmi\pi$, so that (\ref{alevencon}) is satisfied.

There remains the condition (\ref{fcon2}). We must substitute the first of (\ref{fPifX}),  (\ref{fXsol}), (\ref{fqsol}), (\ref{hq0}) and (\ref{hX}). This gives an integral relation that  is again easy to evaluate for coupling functions for which $\alpha^2(\omega)$ is an even function of $\omega$ and $\alpha^2(\omega)/\omega>0$ except possibly at $\omega=0$. Evaluation of the integral by closing the countour in the upper-half plane shows that the integral relation holds without restrictions on the coupling function beyond those already derived from (\ref{fcon1}). This analysis completes the proof that the Hamiltonian is diagonalizable for the class of coupling functions (\ref{alclass}) if and only if  (\ref{alcondiag}) holds.


\end{document}